\documentclass[twocolumn]{aastex701}

\graphicspath{{./}{figures/}}

\usepackage{threeparttable}
\usepackage{color}
\usepackage{graphicx}
\usepackage{multirow}
\usepackage{booktabs}
\usepackage{amsmath}

\def\blue#1 {{\textcolor{blue}{#1}}\ }

\begin{document}

\title{A Bending in the Size-mass Relation of Star-forming Galaxies across $0.5 < z < 6.0$ at a Critical Stellar Mass of $10^{10}M_\odot$ Revealed by JWST}

\author{Longyue Chen}
\affiliation{School of Astronomy and Space Science, Nanjing University, Nanjing, Jiangsu 210093, China}
\affiliation{Key Laboratory of Modern Astronomy and Astrophysics, Nanjing University, Ministry of Education, Nanjing 210093, China}
\email{lychen@smail.nju.edu.cn}

\author[0000-0002-2504-2421]{Tao Wang}
\affiliation{School of Astronomy and Space Science, Nanjing University, Nanjing, Jiangsu 210093, China}
\affiliation{Key Laboratory of Modern Astronomy and Astrophysics, Nanjing University, Ministry of Education, Nanjing 210093, China}
\email{taowang@nju.edu.cn}

\author{Hanwen Sun}
\affiliation{School of Astronomy and Space Science, Nanjing University, Nanjing, Jiangsu 210093, China}
\affiliation{Key Laboratory of Modern Astronomy and Astrophysics, Nanjing University, Ministry of Education, Nanjing 210093, China}
\email{hanwensun@smail.nju.edu.cn}

\author{Ke Xu}
\affiliation{School of Astronomy and Space Science, Nanjing University, Nanjing, Jiangsu 210093, China}
\affiliation{Key Laboratory of Modern Astronomy and Astrophysics, Nanjing University, Ministry of Education, Nanjing 210093, China}
\email{kexu@smail.nju.edu.cn}

\author{Luwenjia Zhou}
\affiliation{School of Astronomy and Space Science, Nanjing University, Nanjing, Jiangsu 210093, China}
\affiliation{Key Laboratory of Modern Astronomy and Astrophysics, Nanjing University, Ministry of Education, Nanjing 210093, China}
\email{wenjia@nju.edu.cn}

\author{Tiancheng Yang}
\affiliation{School of Astronomy and Space Science, Nanjing University, Nanjing, Jiangsu 210093, China}
\affiliation{Key Laboratory of Modern Astronomy and Astrophysics, Nanjing University, Ministry of Education, Nanjing 210093, China}
\email{652023260012@smail.nju.edu.cn}

\author{Maxime Tarrasse}
\affiliation{Université Paris-Saclay, Université Paris Cité, CEA, CNRS, AIM, 91191 Gif-sur-Yvette, France}
\email{maximetarrasse@gmail.com}

\author{Houjun Mo}
\affiliation{Department of Astronomy, University of Massachusetts, Amherst, MA 01003-9305, USA}
\email{hjmo@umass.edu}

\author{Zhaozhou Li}
\affiliation{School of Astronomy and Space Science, Nanjing University, Nanjing, Jiangsu 210093, China}
\affiliation{Key Laboratory of Modern Astronomy and Astrophysics, Nanjing University, Ministry of Education, Nanjing 210093, China}
\email{zhaozhou.li@nju.edu.cn}

\author{Yangyao Chen}
\affiliation{School of Astronomy and Space Science, Nanjing University, Nanjing, Jiangsu 210093, China}
\affiliation{Key Laboratory of Modern Astronomy and Astrophysics, Nanjing University, Ministry of Education, Nanjing 210093, China}
\email{yangyaochen.astro@foxmail.com}

\author{Avishai Dekel}
\affiliation{Center for Astrophysics and Planetary Science, Racah Institute of Physics, The Hebrew University, Jerusalem, 91904, Israel}
\affiliation{Santa Cruz Institute for Particle Physics, University of California, Santa Cruz, CA 95064, USA}
\email{avishai.dekel@mail.huji.ac.il}

\author{Emanuele Daddi}
\affiliation{Université Paris-Saclay, Université Paris Cité, CEA, CNRS, AIM, 91191 Gif-sur-Yvette, France}
\email{emanuele.daddi@cea.fr}

\author{Xuheng Ding}
\affiliation{School of Physics and Technology, Wuhan University, Wuhan 430072, China}
\email{dingxh@whu.edu.cn}

\author{Mauro Giavalisco}
\affiliation{University of Massachusetts Amherst, 710 North Pleasant Street, Amherst, MA 01003-9305, USA}
\email{mauro@umass.edu}

\author{David Elbaz}
\affiliation{Université Paris-Saclay, Université Paris Cité, CEA, CNRS, AIM, 91191 Gif-sur-Yvette, France}
\email{elbaz5439@gmail.com}

\correspondingauthor{Tao Wang}
\email{taowang@nju.edu.cn}

\begin{abstract}
We investigate the rest-frame optical size--stellar mass relation of galaxies at $0.5<z<6.0$ using deep JWST/NIRCam and MIRI imaging from the PRIMER survey. We find that star-forming galaxies (SFGs) exhibit a broken power-law relation at all redshifts, with a nearly constant pivot mass ($M_{\rm p}$) of $\sim 10^{10} M_\odot$, and a slope flattening above $M_{\rm p}$.
This highlights the prevalence of a population of compact, massive SFGs that was underrepresented in previous studies. The size distribution of quiescent galaxies (QGs) is well described by a mixture power-law model, with a pivot mass that increases from $M_{\rm p} \sim 10^{10.0} M_\odot$ at $z =0.75$ to $M_{\rm p} \sim 10^{10.5} M_\odot$ at $z = 2.6$, suggesting that the minimum halo mass required to quench high-mass galaxies increases with redshift. The bending in the size--mass relation of SFGs supports two distinct size growth modes. At $M_{\star} < M_{\rm p}$, size growth is closely coupled to halo growth, while at $M_{\star} > M_{\rm p}$, an increasing fraction of SFGs decouple from halo growth and become compact, likely associated with rapid bulge (and black hole) growth in $M_{\rm h} \gtrsim 10^{12} M_{\odot}$ halos. These compact SFGs are promising progenitors of massive QGs, as evidenced by their similar masses, surface brightness profiles, and morphologies. Their high number densities can account for the observed buildup of massive QGs at $z > 2$, suggesting that the compaction pathway, rather than major mergers of extended SFGs, dominates the formation of high-z massive QGs.
\end{abstract}

\section{Introduction} 
\label{sec:intro}

The relationship between galaxy sizes and stellar masses provides crucial insights into the evolutionary histories of galaxies and the properties of their dark matter halos \citep{2013ApJ...764L..31K,2019ApJ...872L..13M,2019MNRAS.488.4801J}. Observational studies have identified systematic correlations between galaxy sizes and fundamental parameters, including stellar masses, star-formation rates, and redshifts. Because star-forming galaxies (SFGs) and quiescent galaxies (QGs) occupy distinct evolutionary stages, their rest-frame optical size--mass relations are typically modeled independently, employing a single power-law model (e.g., \citealp{2003MNRAS.343..978S,2021ApJ...921...38K,2014ApJ...788...28V,2019ApJ...872L..13M,2024ApJ...972..134M})
. Observational studies up to $z \sim 3$ indicate that both populations exhibit a positive size-mass relation, with SFGs displaying a shallower slope than QGs. 

While the size-mass relations at $z < 3$ have been widely studied, the investigation of high-redshift relations has historically faced observational constraints. A critical factor in size-mass relation studies is the accurate determination of three fundamental parameters: size, stellar mass, and redshift. While galaxy sizes are derived directly from surface brightness profile analysis, determinations of stellar masses and redshifts depend on spectral energy distribution (SED) fitting. Although previous-generation space telescopes—most notably the \textit{Hubble Space Telescope (HST)}—provided rest-frame optical imaging with spatial resolutions of $\sim$1 kpc, such observations were achievable only up to $z \sim 3$. At higher redshifts, the analysis based on \textit{HST} data is naturally limited by the wavelength coverage of its reddest band (F160W), which no longer probes the rest-frame optical regime. 
In addition, the decreasing sensitivity and angular resolution at high redshift make accurate size measurements more difficult. As a result, \textit{HST}-based studies at high redshift are typically focused on the most massive and luminous galaxies, while lower-mass systems become increasingly incomplete.

The advent of the \textit{JWST} has transformed studies of high-redshift galaxies by providing unprecedented angular resolution and sensitivity in the near-infrared. Observations with the Near Infrared Camera (NIRCam) now allow detailed measurements of rest-frame optical morphology for galaxies up to $z \sim 6$, enabling robust investigations of the size--mass relation in the early Universe. In addition, the Mid-Infrared Instrument (MIRI) extends the wavelength coverage beyond rest-frame $1 \,\mu\mathrm{m}$, improving the robustness of stellar mass and redshift estimates, especially for galaxies at $z > 5$ \citep{2025ApJ...988L..35W}. Nonetheless, many early JWST-based studies rely exclusively on NIRCam data, which can introduce systematic uncertainties into size--mass measurements due to the lack of sufficient long-wavelength coverage needed to robustly constrain stellar masses and photometric redshifts \citep{2025ApJ...988L..35W,Sun2026APJL}. Reported findings remain diverse: some studies, based on both simulations and observations, find a positive correlation between galaxy size and stellar mass persisting to $z > 3$ \citep{2024ApJ...962..176W,2024arXiv241016354A}, whereas others suggest flattened or even negative slopes for SFGs with $M_* > 10^{10}M_{\odot}$ \citep{2022MNRAS.514.1921R,2024MNRAS.534.1433S,2024MNRAS.527.6110O,2024MNRAS.533.3724V}. These discrepancies likely reflect limitations of current analyses, including small sample sizes and systematic uncertainties in stellar mass and photometric redshift estimates. Therefore, a precise determination of the size--mass relation based on statistically complete samples and robust parameter estimates is urgently needed for advancing our understanding of galaxy evolution at high redshift.

In this paper, we present the size--mass relations for galaxies over the redshift range $0.5 < z < 6.0$, using both JWST/NIRCam and MIRI data from the Public Release Imaging for Extragalactic Research (PRIMER; GO-1837; PI: J. Dunlop; \citealt{Dunlop2021JWST}). The deep and wide \textit{JWST} observations enable the construction of a statistically robust, mass-complete galaxy sample, reaching stellar masses as low as $\sim 10^{7.9}M_{\odot}$ at $z = 0.5$ and $\sim 10^{8.9}M_{\odot}$ at $z = 5.0$. This coverage allows accurate measurements of the size--mass relation for both SFGs and QGs. We find that SFGs and QGs are better described by broken power-law and mixture models, respectively, which more appropriately characterize their size distributions.

This paper is organized as follows. In Section~\ref{sec:data}, we describe the data set and the SED-fitting procedure employed in this work. Section~\ref{sec:methods} details our methodology, including galaxy size measurements and sample selection criteria. In Section~\ref{sec:results}, we conduct power-law fits of the size--mass relations, and present the derived relations. Section~\ref{sec:discussion} summarizes our main findings and discusses the physical origin of the observed broken power-law behavior. Throughout the paper we adopt a cosmology with $\Omega_m = 0.3, \Omega_\Lambda = 0.7, H_0 = 70\,\mathrm{km\,s^{-1}\,Mpc^{-1}}$. All magnitudes are given in the AB system \citep{1983ApJ...266..713O}, and stellar masses are computed assuming a Kroupa initial mass function \citep{2001MNRAS.322..231K}.

\section{Data} 
\label{sec:data}

\subsection{JWST NIRCam/MIRI Imaging} 

We utilize data from PRIMER \citep{Dunlop2021JWST}, one of the largest public treasury programs conducted with \textit{JWST}. 
The survey includes deep observations covering the CANDELS--COSMOS and CANDELS--UDS fields, obtained in 8 bands with NIRCam (F090W, F115W, F150W, F200W, F277W, F356W, F444W and F410M) and 2 bands with MIRI (F770W and F1800W). To further improve the image quality and wavelength coverage, we also incorporate additional public data obtained with 
\textit{JWST}, \textit{HST}, and ground-based telescopes covering parts of the same fields.  \textit{JWST} data from additional surveys add increased depth and on-sky coverage while HST and ground-based data add wavelength coverage, improving the accuracy of photometric redshifts by $\sim 40\%$, while reducing the outlier fraction by $\sim 60\%$. Further details on the data products and the improvements enabled by the inclusion of these ancillary data are presented in \citet{Sun2026APJL} and \citet{2025ApJ...988L..35W}.

\subsection{Classification of galaxies and Estimation of Photometric Redshifts, Rest-Frame Colors, and Stellar Masses} 

Photometric redshifts are calculated using \texttt{EAzY} \citep{2008ApJ...686.1503B}, incorporating MIRI photometry when available. The \texttt{sfhz.blue\_13} template is applied alongside a 5\% systematic uncertainty floor to the photometric fluxes. Stellar masses and rest-frame $U - V$ and $V - J$ colors are derived through SED fitting with \texttt{BAGPIPES} \citep{2018MNRAS.480.4379C}. The maximum-likelihood redshift from \texttt{EAzY}, also including MIRI photometry when available, is adopted for SED fitting. 

For SED fitting, we use the stellar population synthesis model from the 2016 version of \citet{2003MNRAS.344.1000B} with ages in the range $[0.03, 15] \mathrm{Gyr}$ and metallicity $\log Z/Z_{\odot} = [0.01, 2.5]$. A delayed exponentially declining star formation history with timescales $\tau = [0.01, 10] \, \mathrm{Gyr}$ is adopted, along with dust models from \citet{2000ApJ...533..682C} applied separately to young and old stellar populations, with $A_V$ in the range $[0, 5]$. Nebular emission is included, constructed following \citet{2017ApJ...840...44B}, with an ionization parameter $\log U = [-5, -2]$. 

\citet{2025ApJ...988L..35W} showed that, in the absence of MIRI photometry, the stellar masses of massive galaxies at $z > 5$ are systematically overestimated. Following their results, we apply a correction to the SED-derived stellar masses of galaxies with $M_{\star} > 10^{10}\,M_{\odot}$ at $z > 5$ that lack MIRI coverage, reducing the masses by an average of 0.04 dex at $5.0 < z < 5.5$ and 0.18 dex at $5.5 < z < 6.0$.

\begin{figure*}
 \begin{center}
  \includegraphics[width=\linewidth]{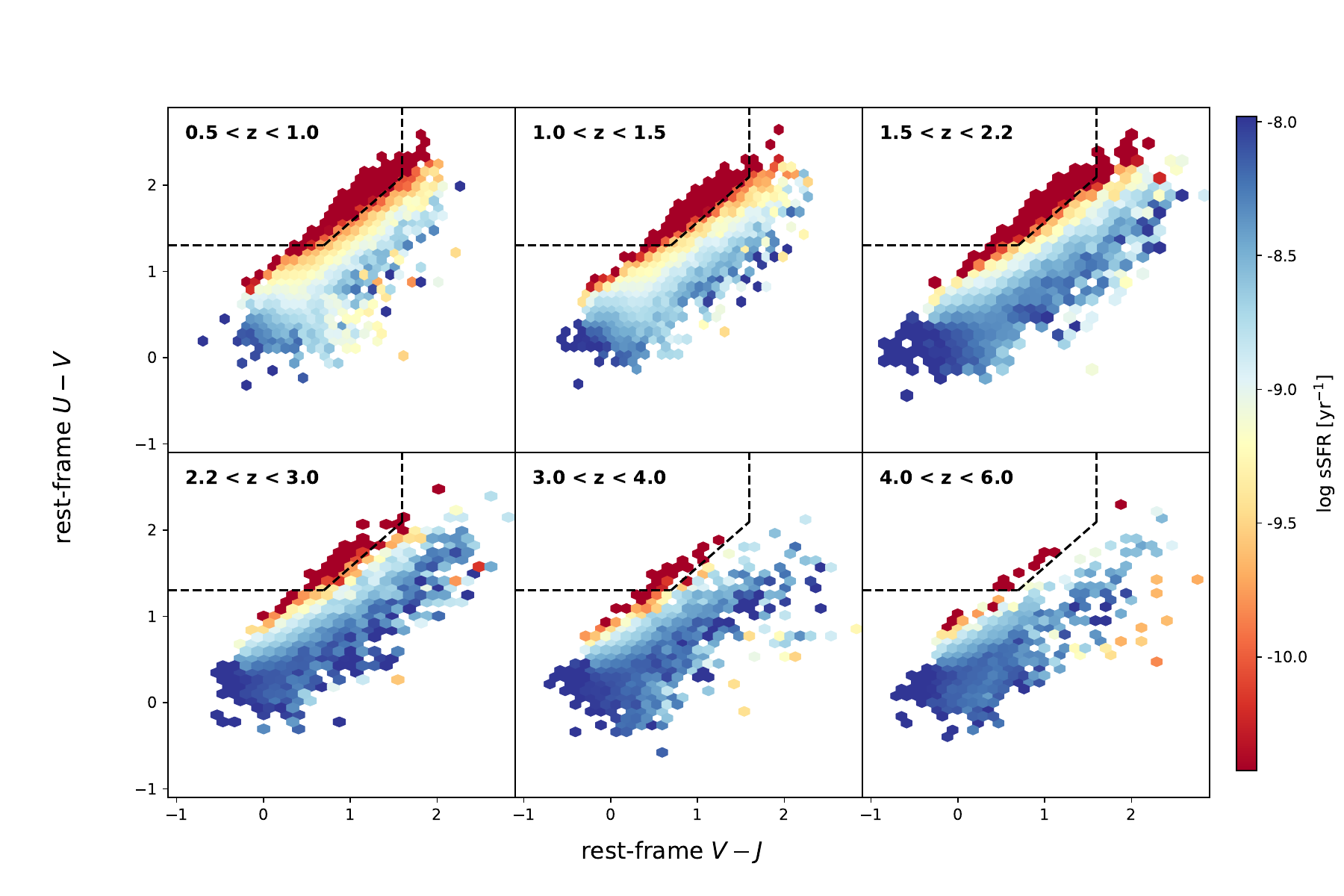}
 \end{center}
\caption{Rest-frame $U - V$ versus $V - J$ color–color diagrams for galaxies in the redshift range $0.5 < z < 6.0$, shown in six redshift bins. The distribution is shown using hexagonal binning, with each bin encoding the mean specific star formation rates (sSFR) of galaxies contributing to that bin. The black dashed lines denote the UVJ selection criterion from \citet{Williams2009ApJ} (see Appendix~\ref{appendix:uvj}).}
\label{fig:ssfr} 
\end{figure*}

Figure~\ref{fig:ssfr} shows the UVJ color-color diagram for all the galaxies in our sample, color-coded by their specific star formation rate (sSFR). While many previous studies have used a non-evolving UVJ color-color criterion to separate the two populations, more recent studies have shown that there is not a universal UVJ criterion applicable to all redshifts~\citep{Merlin2018MNRAS,Belli2019ApJ,2025arXiv251012235Y,Zhang2026ApJ}, as also revealed in Figure~\ref{fig:ssfr}. As a result, we adopt a widely applied sSFR selection criterion \citep[e.g.][]{Pacifici2016ApJ,Singh2025A&A} to identify QGs, as specified by the following conditions:

\begin{equation}
\mathrm{sSFR}_{100} < \frac{0.2}{t_{\mathrm{age}}(z)}
\label{ssfr_criteria}
\end{equation}

where $\mathrm{sSFR}_{100}$ denotes the average specific star formation rate over the most recent 100\,Myr. In this way, a direct comparison can also be made with simulations, in which typically  only sSFR measurements are available. Despite the caveats of using UVJ classification, we also show the size--mass relation results in Appendix~\ref{appendix:uvj} for different populations when using the UVJ classification schemes following \citet{Williams2009ApJ} to facilitate comparison with previous studies. We find that our main conclusions remain unchanged when using UVJ colors to separate QGs from SFGs.

\section{Methods} 
\label{sec:methods}
\subsection{Size Determination} 
\begin{figure*}
 \begin{center}
  \includegraphics[width=\linewidth]{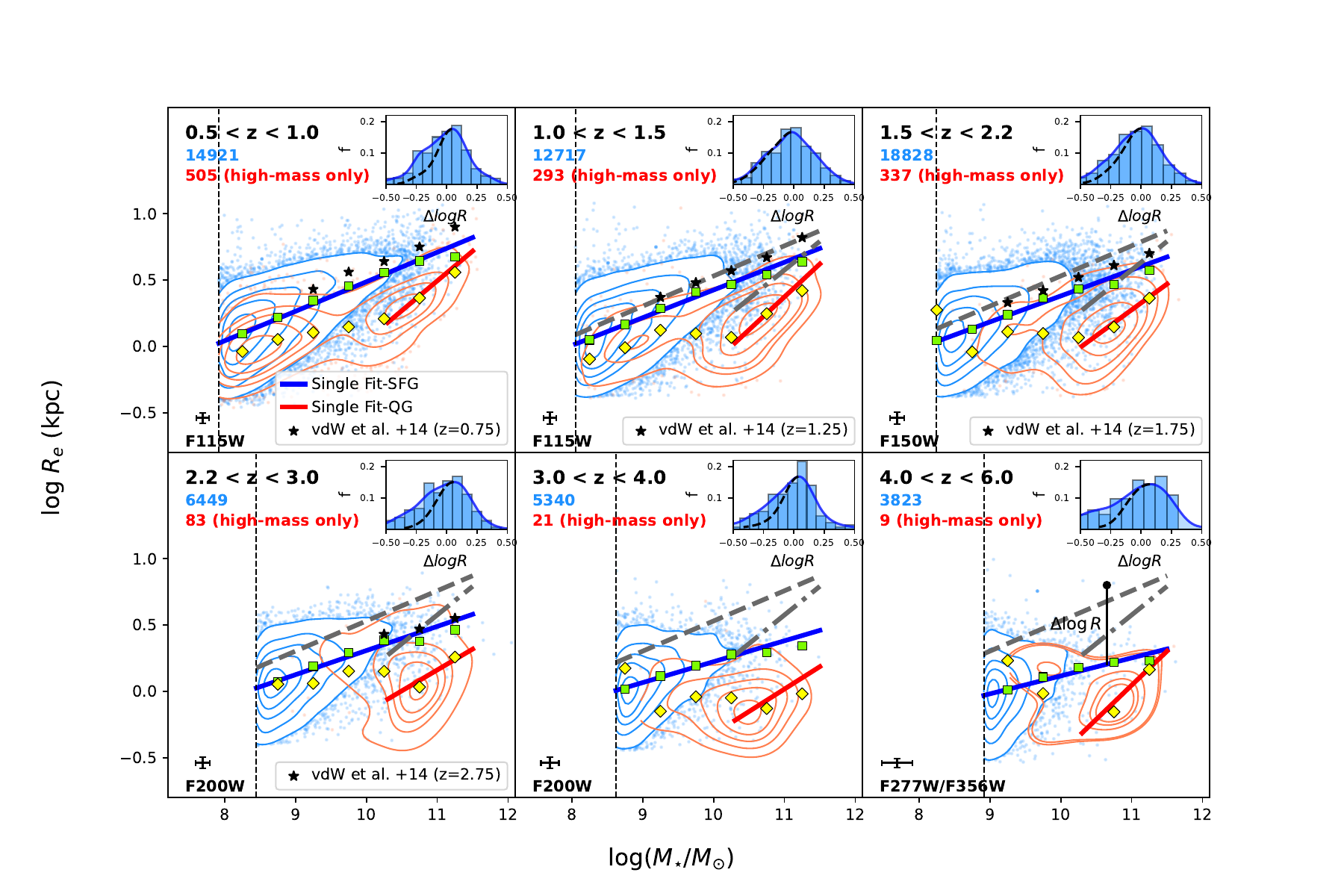}
 \end{center}
\caption{Size--mass relations fitted by a single power-law model for SFGs and high-mass QGs ($M_{\star} > 10^{10.3}M_\odot$). Blue and red contours show the kernel density estimation (KDE)-derived density distributions of SFGs and QGs, with five levels marking the 10th, 30th, 50th, 70th and 90th percentiles. Individual outliers are plotted as blue and red points. Green squares and yellow diamonds indicate median sizes in 0.5-dex mass bins for SFGs and QGs, respectively. Solid lines show the best-fit relations, while dashed (SFGs) and dash–dotted (QGs) reproduce the result at $0.5 < z < 1.0$ for reference. The vertical dashed line marks the mass completeness limit. The bands used for size measurements are labeled in each panel. Specifically, the F277W band is adopted for $4.0 < z < 5.0$ galaxies and the F356W band is used for $5.0 < z < 6.0$ galaxies. Blue and red numbers in the top right indicate the number of SFGs and QGs in each redshift bin. Black stars show the medians of SFGs from \citet{2014ApJ...788...28V}. Insets show the histogram of $\Delta \log R$ distribution for massive SFGs ($M_{\star} > 10^{10.3}M_\odot$) with blue KDE overlaid. The black dashed curve shows a KDE mirrored from the right-hand side of the peak to the left, providing a visual reference for asymmetry. The last panel highlights an example galaxy to illustrate the definition of $\Delta \log R$, where $\Delta \log R$ is defined as the vertical offset from the best-fit single power-law relation.}\label{fig:single} 
\end{figure*}

We use \texttt{galight} \citep{2020ApJ...888...37D} to model each galaxy’s surface brightness profile with a single‐component S\'{e}rsic model. For each field and band, we construct the empirical point spread functions (PSFs) by stacking $\sim20$ bright and unsaturated stars, since the simulated PSFs from \texttt{WebbPSF} \citep{2014SPIE.9143E..3XP} can deviate from empirical PSFs at small angular scales \citep{2024ApJ...974..135J}. During the fitting process, we add a point‐source component whenever it reduces the residual-flux fraction (RFF; see Equation~\ref{equation:RFF}) by more than 10\% and contributes over 10\% of the total galaxy flux. In order to improve the fitting accuracy and efficiency, especially for galaxies with low signal-to-noise ratio (S/N), we first fit all galaxies using a detection image constructed as a weighted stack of the F277W, F356W, F410M, and F444W images. The stacking is performed on a pixel-by-pixel basis, such that only bands with valid coverage contribute to each pixel. The resulting fits are then used to obtain prior parameter estimates. We then fix the galaxy center from these priors and adopt the remaining parameters as initial guesses for a single‐band fit in the filter closest to the rest‐frame 5000\,\AA (as indicated in the lower left of each panel of Figure~\ref{fig:single}). Two consecutive rounds of Particle Swarm Optimization (PSO) and one round of Markov Chain Monte Carlo (MCMC) are employed throughout all the fitting processes. Neighboring sources within each cutout are treated based on their segmentation ellipses. For each object, we expand the semi-major and semi-minor axes of the segmentation ellipse by a factor of 2 and test for overlap with the correspondingly expanded ellipse of the target galaxy. Sources whose expanded ellipses intersect with that of the target are fit simultaneously, while the others are masked.

\begin{figure*}
\centering
\includegraphics[width=\linewidth]{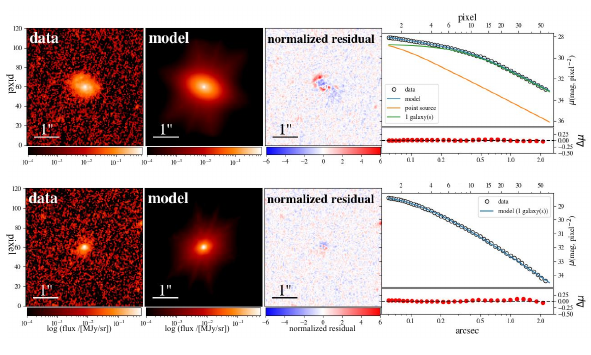}
\caption{Examples of \texttt{galight} fitting results. In each row, the images show original data, model, normalized residuals, and 1D surface brightness profile from left to right. The upper panel shows one galaxy fitted by both a single S\'{e}rsic model and a point source model, with $z = 2.2$, $M_{\star} = 10^{10.5} M_{\odot}$ and $R_{e} = 2.7 \, \text{kpc}$. The lower panel shows another galaxy fitted by only a single S\'{e}rsic model, with $z = 3.5$, $M_{\star} = 10^{10.6} M_{\odot}$ and $R_{e} = 0.9 \, \text{kpc}$.}\label{fig:galight_example} 
\end{figure*}

Examples of fitting results are shown in Figure~\ref{fig:galight_example}. We quantify galaxy sizes by their effective radius ($R_e$), defined as the semi-major axis of the ellipse that encloses half of the total flux of the best-fit S\'{e}rsic model.

To assess fitting quality, we calculate the residual flux
fraction (RFF) for each galaxy, which quantifies the signal in the residual image that cannot be attributed to background fluctuations \citep{2012MNRAS.419.2703H}. Following \citet{2016MNRAS.461.2728M}, RFF is defined as
\begin{equation}
RFF = \frac{\sum_{j,k \in A} \left| I_{j,k} - I_{j,k}^{\text{galight}} \right| - 0.8 N \langle \sigma_B \rangle}{\text{FLUX\_AUTO}}
\label{equation:RFF}
\end{equation}
where $I_{j,k}$ and $I_{j,k}^{\text{galight}}$ represent the fluxes in the $j$-th and $k$-th pixels of the science and model images estimated by \texttt{galight}, respectively. \(\langle \sigma_B \rangle\) is the mean value of the sky background, and FLUX\_AUTO is the flux of the galaxy calculated by \texttt{Source Extractor}. The value of $N$ is the number of pixels within the Kron ellipse. The factor of 0.8 in the numerator ensures that the average value of the RFF vanishes to zero when a Gaussian error image with constant variance is encountered. We also assess the fitting quality by computing the chi-squared value of each fit and find that this metric has negligible impact on the results when compared to a limit on the RFF.

\subsection{Sample Selection} 
Mass completeness limits are determined at the median redshift of each bin following \citet{Sun2026APJL}, and this single completeness limit is adopted for all galaxies within each bin. The limits range from $M_{\star}=10^{7.9}M_{\odot}$ at $0.5<z<1.0$ to $M_{\star}=10^{8.9}M_{\odot}$ at $4.0<z<6.0$. As demonstrated in Appendix~\ref{appendix:galight}, \texttt{galight} reliably recovers galaxy sizes for sources with intrinsic magnitudes $\leq 27.5$ and fitted effective radii $R_e \geq 0.05\,\arcsec$. Therefore, we adopt a magnitude limit of 27.5 across all redshifts. To ensure robust detection and reliable structural parameter estimate, we require a S/N of galaxies greater than 5 in the band used for size measurements. 

Our mass-, magnitude-, and S/N-limited sample comprises 74813 sources (70137 SFGs and 3580 QGs). We first remove 4114 sources with fitted $|RFF| > 0.5$, $R_e \leq 0.05\, \arcsec$ or $R_e$ larger than the radius of image cutouts ($R_e \geq 1.5\, \arcsec$) due to their low confidence. Then, we manually verified the SED results and multi-band images for all sources with a S\'{e}rsic index $n$ that reaches the boundaries of the bounds ($0.3 \leq n \leq 9$), or with an $R_e$ that is too small or too large ($R_e \leq 0.6 \, \text{kpc}$ or $R_e \geq 4.0 \, \text{kpc}$), removing 2245 sources with errors in size determination after the check. We also exclude 2414 SFGs in the CANDELS--COSMOS field at $2.2 < z < 2.8$ to mitigate the impact of the known supercluster structure \citep{Cucciati2018}, as SFGs in overdense environments are found to follow size--mass relations that differ from those of their field counterparts \citep{Xu_2023} (see Appendix~\ref{appendix:cosmos_error} for further discussion). The final sample contains 64908 galaxies (61667 SFGs and 3241 QGs), including 9204 galaxies at $z \geq 3.0$. For our galaxy sample, the comparison between photometric and spectroscopic redshifts yields a normalized median absolute deviation of $\sigma_{\rm NMAD} = 0.016$. The fraction of catastrophic outliers is $f_{\rm outlier} = 2.2\%$, demonstrating the overall reliability of the photometric redshifts \citep{Sun2026APJL}.

\begin{deluxetable*}{lcccccc}
\tablecaption{Best-fit results of the size--mass relations of the form $R_e/{\rm kpc} = A\,(M_{\star}/5\times10^{10}M_\odot)^{k}$ (Equation~\ref{eq:single})
for SFGs and high-mass QGs ($M_{\star} > 10^{10.3}M_\odot$). 
\label{tab:table_single}}
\tablehead{
  \colhead{\multirow{2}{*}{Redshift}} & \multicolumn{3}{c}{Star-forming galaxies} & \multicolumn{3}{c}{Quiescent galaxies} \\
  \cmidrule(lr){2-4} \cmidrule(lr){5-7}
  & \colhead{$k$} & \colhead{$\log(A)$} & \colhead{$\sigma\log(R_{\rm eff})$} & \colhead{$k$} & \colhead{$\log(A)$} & \colhead{$\sigma\log(R_{\rm eff})$}
}

\startdata
$0.5<z<1.0$ & $0.22_{-0.01}^{+0.01}$ & $0.64_{-0.01}^{+0.01}$ & $0.20_{-0.01}^{+0.01}$ & $0.45_{-0.02}^{+0.03}$ & $0.36_{-0.01}^{+0.01}$ & $0.16_{-0.01}^{+0.01}$ \\
$1.0<z<1.5$ & $0.21_{+0.01}^{-0.01}$ & $0.57_{+0.01}^{-0.01}$ & $0.20_{+0.01}^{-0.01}$ & $0.50_{-0.04}^{+0.04}$ & $0.22_{+0.01}^{-0.01}$ & $0.17_{+0.01}^{-0.01}$ \\
$1.5<z<2.2$ & $0.20_{-0.01}^{+0.01}$ & $0.52_{-0.01}^{+0.01}$ & $0.20_{-0.01}^{+0.01}$ & $0.39_{-0.04}^{+0.04}$  & $0.16_{-0.01}^{+0.01}$ & $0.20_{-0.01}^{+0.01}$ \\
$2.2<z<3.0$ & $0.18_{-0.01}^{+0.01}$ & $0.43_{-0.01}^{+0.01}$ & $0.19_{-0.01}^{+0.01}$ & $0.31_{-0.09}^{+0.09}$ & $0.07_{-0.02}^{+0.02}$ & $0.22_{-0.02}^{+0.01}$ \\
$3.0<z<4.0$ & $0.16_{-0.01}^{+0.01}$ & $0.33_{-0.01}^{+0.01}$ & $0.19_{-0.01}^{+0.01}$ & $0.34_{-0.14}^{+0.21}$ & $-0.08_{-0.03}^{+0.05}$ & $0.18_{-0.04}^{+0.02}$\\
$4.0<z<6.0$ & $0.14_{-0.01}^{+0.01}$ & $0.21_{-0.02}^{+0.02}$ & $0.20_{-0.01}^{+0.01}$ & $0.51_{-0.14}^{+0.17}$ & $-0.11_{-0.08}^{+0.06}$ & $0.13_{-0.05}^{+0.01}$\\
\enddata
\tablecomments{For QGs at $z > 3$, the uncertainties in the fitted parameters increase owing to the small number of samples.}
\end{deluxetable*}

\section{\texorpdfstring{The Size--Mass Relation at $0.5 < \lowercase{z} < 6.0$}{The Size--Mass Relation at 0.5 < z < 6.0}}
\label{sec:results}

\subsection{Single Power-law Model fits of Size--Mass Relation} \label{subsec:single_fit}

The high-resolution, large-sample data from \textit{JWST} observations allow us to conduct more detailed studies of galaxy size distributions. Here, we first model the size--mass relation using a single power-law form.

Following \citet{2014ApJ...788...28V} and \citet{2021ApJ...921...38K}, the single power-law function is defined as:
\begin{equation}
R_e(M_{\star})/\text{kpc} = A\left( \frac{M_{\star}}{5\times10^{10}M_\odot} \right)^{k}
\label{eq:single}
\end{equation}
where $k$ is the power-law slope, and $A$ is the effective radius at $M_{\star} = 5\times10^{10}M_\odot$.

We adopt the methods described by \citet{2014ApJ...788...28V} to obtain the best-fit parameters of the model by maximizing the likelihood. As a prior, we assume a log-normal distribution in galaxy size, $N(\log r, \sigma_{\log r})$, where $r(m_{\star})/\text{kpc} = A \left({M_{\star}}/{5\times10^{10}M_\odot} \right)^{k}$. If the measured $R_e$ also has a Gaussian, 1-$\sigma$ uncertainty of $\delta \log R_e$, the probability for observed QGs ($P_{QG}$) and SFGs ($P_{SFG}$) is given by:
\begin{equation}
P = \langle \mathcal{N}(\log R_{\mathrm{eff}}, \delta \log R_{\mathrm{eff}}),\, 
\mathcal{N}(\log r(m_{\star}), \sigma_{\log r}) \rangle
\end{equation}
The final likelihood adopted in the fitting is given by $\mathcal{L} = \sum \ln P_{SFG}$ or $\mathcal{L} = \sum \ln P_{QG}$. The likelihood is maximized within each redshift bin to determine the optimal values including $k$, $A$, and intrinsic scatter $\sigma$ for each single power-law model.

The size as a function of stellar mass at $0.5 < z < 6.0$ for SFGs and high-mass QGs is shown in Figure~\ref{fig:single}. Detailed values of fitting parameters are tabulated in the Table~\ref{tab:table_single}. When using the single power-law models for fitting, the size exhibits a statistically significant positive correlation with the stellar mass of both types of galaxies across all redshifts. For SFGs, the slopes of the size--mass relations decrease with increasing redshift. High-mass QGs follow a size--mass relation with a steeper slope and lower intercept than SFGs.

\subsection{Broken Power-law and Mixture Model fits of Size--Mass Relation}\label{subsec:broken_fit}

Although the size--mass relations fitted with a single power-law model can roughly reproduce the overall size distributions, the medians show clear deviations, particularly at the high-mass end for SFGs and the low-mass end for QGs. For SFGs, this mainly reflects the sample being dominated by low-mass galaxies, which biases the fitted slope, leading to an underestimation of the sizes of massive galaxies. The insets in Figure~\ref{fig:single} highlight the asymmetric distribution of massive galaxies relative to the best-fit single power-law model, indicating an excess of massive and compact SFGs. For QGs, low-mass galaxies are known to follow a different size--mass relation from massive ones, likely linked to variations in quenching mechanisms and evolutionary pathways \citep{2003MNRAS.343..978S,Whitaker2017ApJ,2021MNRAS.505..172L,2021MNRAS.506..928N,Cutler2022ApJ,2024ApJ...967L..23C,Hamadouche2025MNRAS}.
To better capture the distinct behaviors of low- and high-mass galaxies, we therefore adopt a broken power-law model and a mixture power-law model to fit the size--mass relation.

For SFGs, we adopt the smoothly broken power-law function following \citet{2021ApJ...921...38K}, which is defined as:
\begin{equation}
R_e(M_{\star})/\text{kpc} = r_p\left( \frac{M_{\star}}{M_p} \right)^{\alpha}\left[\frac{1}{2}\left\{1 + \left( \frac{M_{\star}}{M_p}\right)^\delta \right\} \right]^{\frac{\beta - \alpha}{\delta}}
\label{eq:broken_sfg}
\end{equation}
where $M_p$ is the pivot stellar mass, $r_p$ is the effective radius at the pivot stellar mass, $\alpha$ is the slope at the low-mass end, $\beta$ is the slope at the high-mass end, and $\delta$ is the smoothing factor, which is set to 6 to reduce the degeneracy between $\delta$ and the slopes. We then use the same methods as Section~\ref{subsec:single_fit} to derive the best-fit parameters, including $\alpha$, $\beta$, $M_p$, $r_p$ and $\sigma$ for each broken power-law model.

For QGs, low- and high-mass systems appear to form two distinct populations, making Equation~\ref{eq:broken_sfg} insufficient to describe the size distribution (see Figure~\ref{appendix:fig_qg} and Table~\ref{appendix:table_qg}). We therefore model the size–mass relation using a mixture of two power-law components. The mixture model is defined as:
\begin{equation}
\begin{aligned}
\log R_{\mathrm{e}}/\text{kpc} =\;&
\left[1 - f(M_\star)\right]
\left[a_1 (\log M_\star - \log m_0) + b_1\right] \\
&+
f(M_\star)
\left[a_2 (\log M_\star - \log m_0) + b_2\right],
\end{aligned}
\label{eq:broken_qg}
\end{equation}

where the transition fraction is described by a logistic function,

\begin{equation}
f(M_\star) =
\frac{1}{1 + \exp \left[-(\log M_\star - \log M_p)/W \right]}.
\end{equation}

Here $a_1$ and $a_2$ denote the slopes of the low- and high-mass power-law components, respectively.
$b_1$ and $b_2$ are the intercepts evaluated at the pivot mass $m_0$, which is set to $5\times10^{10}M_\odot$. The parameter $M_p$ corresponds to the pivot stellar mass at which the transition between the two QG populations occurs. The parameter $W$ controls the width of this transition in stellar mass. We also use the methods as~\ref{subsec:single_fit} to derive the best-fit parameters, including $a_1$, $b_1$, $a_2$, $b_2$, $M_p$ and $W$ for each mixture model.

\begin{figure*}
\centering
\includegraphics[width=\textwidth]{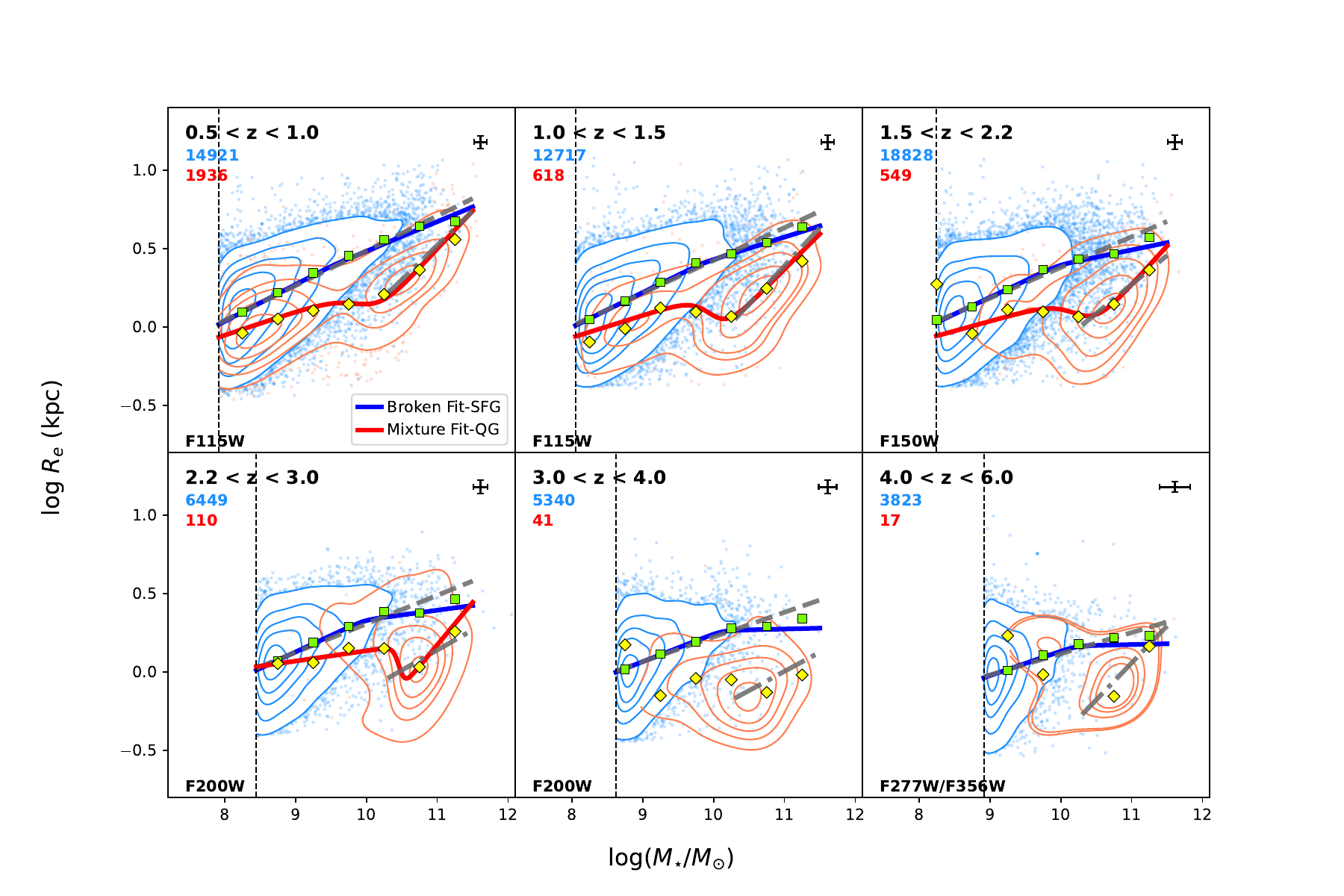}
\caption{Size--mass relations fitted by a broken power-law model for SFGs at $0.5 < z < 6.0$ and a mixture model for QGs at $0.5 < z < 4.0$ (same symbols as in Figure~\ref{fig:single}). At $z > 4.0$, the lack of low-mass QGs results in an unconstrained mixture model fit so the mixture model is not applicable. Grey dashed lines represent the relations fitted by single power-law shown in Figure~\ref{fig:single} for comparison.}\label{fig:broken} 
\end{figure*}

The size--mass relations for SFGs and QGs fitted with broken and mixture power-law models are shown in Figure~\ref{fig:broken}, and the corresponding best-fit parameters are listed in Tables~\ref{tab:table_broken_sfg} and~\ref{tab:table_broken_qg}. For SFGs, the low-mass end of the broken power-law relation closely matches that obtained with a single power-law fit, while the high-mass end becomes noticeably flatter. At $z > 2.2$, SFGs with $M_{\star} > M_p$ exhibit minor size dependence on stellar mass. For QGs, the mixture model fits reveal shallower slopes at the low-mass end than at the high-mass end, consistent with previous findings \citep{2003MNRAS.343..978S,2021MNRAS.506..928N,2024ApJ...967L..23C}. 

To assess whether the improvement achieved by the broken power-law model justifies its additional degrees of freedom for SFGs, we compare the Bayesian Information Criterion (BIC; \citealt{1978AnSta...6..461S}) between the two models. The BIC is defined as
\begin{equation}
    \mathrm{BIC} = k_{\rm BIC} \ln(N_{\rm BIC}) - 2 \ln(\hat{L}),
\end{equation}
where \(k_{\rm BIC}\) is the number of free parameters in the model, \(N_{\rm BIC}\) is the number of data points used for the BIC calculation, and \(\hat{L}\) is the maximum likelihood of the model. This criterion penalizes models with more free parameters, thereby favoring the simplest model that adequately describes the data. A difference of 
\(\Delta\mathrm{BIC} = \mathrm{BIC}_{\rm single} - \mathrm{BIC}_{\rm broken} > 0\) 
indicates that the broken power-law provides a statistically significant improvement over the single power-law model.

As shown in Table~\ref{tab:table_broken_sfg}, for SFGs at \(1.0 < z < 4.0\), the broken power-law consistently yields \(\Delta \mathrm{BIC} > 0\), demonstrating that the improvement in the fit is statistically meaningful and supporting the presence of a pronounced flattening in the size--mass relation above the pivot mass. For SFGs at \(4.0 < z < 6.0\), the small sample size, especially for high-mass galaxies, causes the BIC to favor the simpler model (i.e., the single power-law). For SFGs at \(0.5 < z < 1.0\), we suppose that the emergence of more diverse quench mechanisms has reduced the proportion of compact SFGs, making the BIC favor the simpler model (see Section~\ref{sec:discussion} for more discussion). These results indicate that the bending feature is not an artifact of model selection, but instead reflects a genuine size transition in massive SFGs.

\begin{deluxetable*}{lccccccc}
\tablecaption{Best-fit results of the size--mass relations for SFGs of the form $R_e/{\rm kpc} = r_p(M_{\star}/M_p)^{\alpha}\left[0.5\{1 + (M_{\star}/M_p)^{\delta}\}\right]^{(\beta-\alpha)/\delta}$.\label{tab:table_broken_sfg}}
\tablehead{
  \colhead{Redshift} & \colhead{$\alpha$} & \colhead{$\beta$} & \colhead{$\log(M_{\rm p})$} &
  \colhead{$\log(r_{\rm p})$} & \colhead{$\sigma\log(R_{\rm eff})$} & \colhead{$\Delta BIC$}
}
\startdata
$0.5<z<1.0$ & $0.23_{-0.01}^{+0.01}$ & $0.19_{-0.01}^{+0.01}$ & $9.53_{-0.16}^{+0.22}$ & $0.39_{-0.03}^{+0.05}$ & $0.20_{-0.01}^{+0.01}$ & -1.91 \\
$1.0<z<1.5$ & $0.22_{-0.01}^{+0.01}$ & $0.14_{-0.02}^{+0.01}$ & $9.85_{-0.09}^{+0.10}$ & $0.41_{-0.02}^{+0.02}$ & $0.20_{-0.01}^{+0.01}$ & 7.68 \\
$1.5<z<2.2$ & $0.21_{-0.01}^{+0.01}$ & $0.09_{-0.02}^{+0.01}$ & $9.97_{-0.06}^{+0.06}$ & $0.39_{-0.01}^{+0.01}$ & $0.20_{-0.01}^{+0.01}$ & 60.17 \\
$2.2<z<3.0$ & $0.21_{-0.01}^{+0.01}$ & $0.06_{-0.03}^{+0.02}$ & $9.97_{-0.11}^{+0.13}$ & $0.32_{-0.02}^{+0.02}$ & $0.19_{-0.01}^{+0.01}$ & 31.82 \\
$3.0<z<4.0$ & $0.18_{-0.01}^{+0.01}$ & $0.01_{-0.04}^{+0.04}$ & $10.12_{-0.14}^{+0.10}$ & $0.26_{-0.02}^{+0.02}$ & $0.19_{-0.01}^{+0.01}$ & 6.10 \\
$4.0<z<6.0$ & $0.16_{-0.01}^{+0.01}$ & $0.01_{-0.09}^{+0.03}$ & $10.22_{-0.21}^{+0.44}$ & $0.16_{-0.03}^{+0.06}$ & $0.20_{-0.01}^{+0.01}$ & -5.89 \\
\enddata
\end{deluxetable*}

\begin{deluxetable*}{lcccccccc}
\tablecaption{Best-fit results of the size-mass relations for QGs of the form $\log R_{\mathrm{e}} =
\left[1 - f(M_\star)\right]
\left[a_1 (\log M_\star - 10.5) + b_1\right] +
f(M_\star)
\left[a_2 (\log M_\star - 10.5) + b_2\right]$ (Equation~\ref{eq:broken_qg}).}\label{tab:table_broken_qg}
\tablehead{
  \colhead{Redshift} & \colhead{$a_1$} & \colhead{$b_1$} & \colhead{$\sigma\log(R_{\rm eff1})$} & \colhead{$a_2$} & \colhead{$b_2$} & \colhead{$\sigma\log(R_{\rm eff2})$} &
  \colhead{$\log(M_{\rm p})$} & \colhead{$W$}
}
\startdata
$0.5<z<1.0$ & $0.15_{-0.01}^{+0.01}$ & $0.35_{-0.02}^{+0.02}$ & $0.18_{-0.01}^{+0.01}$ & $0.48_{-0.03}^{+0.03}$ & $0.36_{-0.01}^{+0.01}$& $0.16_{-0.01}^{+0.01}$ & $10.05_{-0.06}^{+0.07}$& $0.20_{-0.03}^{+0.02}$\\
$1.0<z<1.5$ & $0.14_{-0.03}^{+0.03}$ & $0.32_{-0.05}^{+0.05}$ & $0.20_{-0.01}^{+0.01}$ & $0.45_{-0.05}^{+0.04}$ & $0.22_{-0.01}^{+0.01}$& $0.17_{-0.01}^{+0.01}$ & $10.01_{-0.11}^{+0.11}$& $0.14_{-0.04}^{+0.02}$\\
$1.5<z<2.2$ & $0.13_{-0.04}^{+0.04}$ & $0.25_{-0.05}^{+0.06}$ & $0.19_{-0.01}^{+0.01}$ & $0.49_{-0.07}^{+0.08}$ & $0.12_{-0.04}^{+0.03}$& $0.17_{-0.02}^{+0.01}$ & $10.33_{-0.16}^{+0.20}$& $0.17_{-0.06}^{+0.03}$\\
$2.2<z<3.0$ & $0.07_{-0.07}^{+0.16}$ & $0.18_{-0.06}^{+0.09}$ & $0.25_{-0.04}^{+0.02}$ & $0.54_{-0.03}^{+0.01}$ & $0.01_{-0.04}^{+0.02}$& $0.19_{-0.03}^{+0.01}$ & $10.47_{-0.04}^{+0.09}$& $0.04_{-0.04}^{+0.04}$\\
\enddata
\end{deluxetable*}

\subsection{Redshift Evolution of the size--mass relation} 
\begin{figure*}
 \begin{center}
  \includegraphics[width=\linewidth]{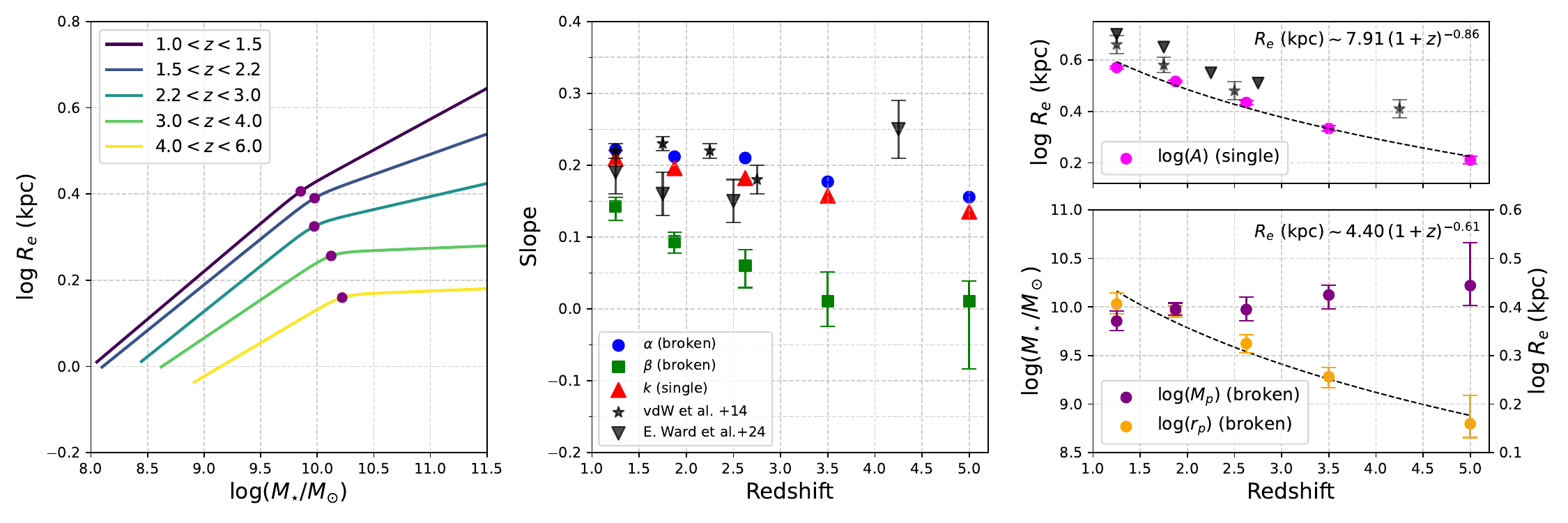}
 \end{center}
\caption{Redshift evolution of the size--mass relations and associated parameters for SFGs at $1.0 < z < 6.0$. {\it Left}: Evolution of the size-mass relations for SFGs. The solid lines represent the best-fit relations using a broken power-law model, with pivots indicated by purple dots. {\it Middle}: Evolution of the parameters from power-law fits, including $k$ (red triangle), $\alpha$ (blue dots) and $\beta$ (green squares) from broken power-law fits. {\it Right}: Evolution of the parameters from power-law fits in logarithmic scale. The top panel shows the intercept (magenta dots) from the single power-law model fits, while the bottom panel shows the pivot mass (purple dots) and pivot $R_e$ (orange dots) from broken power-law model fits. The best-fit single power-law parameters from \citet{2014ApJ...788...28V} (black stars) and \citet{2024ApJ...962..176W} (black inverted triangles) are also shown for comparison.
}\label{fig:error}
\end{figure*}

The left panel of Figure~\ref{fig:error} illustrates the best-fitting broken power-law models at different redshifts for SFGs, while the middle and right panels depict the evolution of key best-fit parameters for both single and broken power-law models for SFGs. For comparison, we also plot the slopes and intercepts fitted by single power-law models reported by \citet{2014ApJ...788...28V} and \citet{2024ApJ...962..176W} in the middle and right-top panels, respectively. The similarity between our single power‐law results and these earlier studies further validates the robustness of our methodology.

We find that the evolutionary trend of the low-mass end slope ($\alpha$) is similar to the slopes obtained from single power-law model fits ($k$) for SFGs, but $\alpha$ is consistently higher than $k$ across all redshifts. We note that $\alpha$ is approximately constant at 0.21 at $z < 3$, but the redshift evolution of the high-mass end slope ($\beta$) shows significant divergence. This parameter evolves from $\beta = 0.19$ at $z = 0.75$ to $\beta = 0.01$ at $z = 5.0$, creating an increasing disparity between $\alpha$ and $\beta$ that expands from 0.04 at $z = 0.75$ to 0.15 at $z = 5.0$. The pivot radius decreases with redshift, declining from $\log r_p = 0.39 \,\mathrm{kpc}$ at $z = 0.75$ to $\log r_p = 0.16 \,\mathrm{kpc}$ at $z = 5.0$. The pivot mass, $M_p$, is approximately constant at $10^{10}M_\odot$ at all redshifts. 

\section{Discussion}
\label{sec:discussion}
Utilizing the high-resolution imaging from \textit{JWST} and the improved SED-fitting results enabled by the inclusion of \textit{MIRI} data, we derive the size--mass relations across $0.5 < z < 6.0$. In particular, we show that the size--mass relations for SFGs are well fitted by a broken power-law model, with a pivot stellar mass of $M_{\star} \sim 10^{10}M_{\odot}$ that persists across the full redshift range considered. Here we compare our results with previous studies and discuss the implications of these findings on the evolutionary path of massive galaxies over cosmic time.

\subsection{Comparison with previous studies}
Our size--mass relations derived from single power-law fits are consistent with previous studies by \citet{2014ApJ...788...28V} and \citet{2024ApJ...962..176W}, as shown in Figures~\ref{fig:single} and~\ref{fig:error}. The relation between $R_e$ and $z$, shown in the upper-right panel of Figure~\ref{fig:error}, is also consistent with previous results \citep{2014ApJ...788...28V,2024ApJ...962..176W,Li2024A&A}. However, for SFGs we identify a shallower relation at the high-mass end that has not been widely emphasized in earlier work.
We attribute this to the limited spatial resolution of \textit{HST} imaging, which can compromise or prevent accurate size measurements of intrinsically compact, high-mass galaxies, either through systematic overestimation of $R_e$ or through contamination from nearby bright sources, thereby suppressing the flattened trend revealed by \textit{JWST}. This bias in the $R_e$ measurement of high-mass galaxies with \textit{HST} imaging is examined further in Appendix~\ref{appendix:compare}.

Our results indicate that a broken power-law model more accurately describes the size--mass relation of SFGs, revealing a slope change (i.e., a bending) at the high-mass end. These results are consistent with some simulations \citep{2023ApJ...946...71C,2024MNRAS.534.1433S}, COSMOS-Web findings \citep{2025arXiv250407185Y}, and other early studies \citep{2017ApJ...840...47B,2024arXiv241214970W,2024MNRAS.527.6110O,2024MNRAS.533.3724V}, all of which report flattened or even inverted size--mass relations at high redshift in the massive regime. Our findings suggest that previous fits of the size--mass relations may have been predominantly governed by low-mass galaxies, potentially underestimating the role and prevalence of compact, massive galaxies. A broken power-law model successfully characterizes the size distribution of both low-mass and high-mass SFGs, offering new insights into the divergent evolutionary processes of SFGs with different masses.

\subsection{Implications on the structural evolution of SFGs over cosmic time}
\label{subsec:Implications on the structural}

The broken power-law behavior of the size-mass relations for SFGs indicates two modes of size growth diverging at a pivot stellar mass of $M_{\star} \sim 10^{10}M_{\odot}$. Below this mass, the sizes of SFGs are well described by a single power-law relation, in which galaxy size increases more sensitively with stellar mass. This relation remains valid across the broad redshift range considered in this study. In contrast, galaxies above the pivot mass exhibit a different growth mode with a shallower and even flat power-law slope. 

The bending of the size--mass relation at the high-mass end is largely driven by a population of massive and compact SFGs emerging above the pivot mass ($> 10^{10.0}M_{\odot}$). However, it is clear that not all massive galaxies above the pivot mass are compact. We classify massive SFGs into two populations based on their offsets from the best-fit single power-law size--mass relation (i.e., $\Delta \log R$, as illustrated in the last panel of Figure~\ref{fig:single}):
\begin{itemize}
    \item {\it{Compact star-forming galaxies (cSFGs)}}: SFGs with $\Delta \log R < -1\sigma$, i.e., galaxies lying more than $1\sigma$ below the best-fit single power-law size--mass relation.
    \item {\it{Extended star-forming galaxies (eSFGs)}}: SFGs with $|\Delta \log R| \leq 1\sigma$, i.e., galaxies consistent with the size--mass relation within $1\sigma$.
\end{itemize}

To investigate the structural connection between eSFGs, cSFGs, and QGs, we construct matched triplets of galaxies within $2.2 < z < 4.0$ and $10^{10.0} M_{\odot} < M_{\star} < 10^{10.8} M_{\odot}$, with the stellar mass of each galaxy in a triplet differing by less than 0.1 dex to minimize mass-difference biases. We then stack their F200W radial surface brightness profiles and fit the stacked profiles with single S\'{e}rsic models, as shown in the right-top panel of Figure~\ref{fig:cSFGs}. The stacked profile of cSFGs lies between those of QGs and eSFGs, reflecting their intermediate $R_e$ and S\'{e}rsic indices.

We examine the axis ratios ($b/a$) and S\'{e}rsic indices for the full sample across all redshifts (right-bottom panel of Figure~\ref{fig:cSFGs} and Figure~\ref{fig:sersic}). cSFGs exhibit $b/a$ distributions similar to QGs and systematically higher than eSFGs, while their S\'{e}rsic indices are typically around $n \sim 2$, reflecting their intermediate, somewhat spheroidal morphology. These trends indicate that cSFGs are not simply SFGs with smaller disks, but instead have more centrally concentrated structures, consistent with their compact sizes and supporting their role as likely progenitors of QGs. Their transitional morphologies further suggest that star formation in cSFGs may be preferentially concentrated toward central regions, which over time can steepen the light profile, producing smaller effective radii ($R_e$) and higher S\'{e}rsic indices.

Our findings on the nearly constant slope for the size--mass relation at $M_{\star} < M_p$ are broadly consistent with previous results, and suggest that the size growth at low masses may be tightly linked to their host halo spin parameter \citep{1998MNRAS.295..319M,2014ApJ...788...28V}, but see \cite{JIANGFZ2019} for a different explanation. 

On the other hand, the bending in the size--mass relation at a critical mass suggests an increasing fraction of SFGs decouple from their halo growth. This bending behavior may mirror trends seen in both the stellar-to-halo mass relation (SHMR; \citealt{2013MNRAS.428.3121M,2019MNRAS.488.3143B}) and the star-formation main sequence (SFMS; \citealt{2022A&A...661L...7D}). The ratio $M_{\star}/M_{\rm halo}$ has been shown to peak at $M_{\rm halo}^{\rm peak} \sim 10^{12} M_{\odot}$, corresponding to galaxies with stellar masses of $M_{\star}^{\rm peak} \sim 10^{10}M_{\odot}$ — a value remarkably close to the observed pivot mass in the size--mass relation. Below this characteristic mass, stellar mass growth generally tracks halo mass growth, and galaxies exhibit steadily increasing star-formation rates with stellar mass. Above the SHMR peak, however, this coupling begins to break down: halo mass continues to grow efficiently, while stellar mass assembly and star formation become progressively less efficient, leading to a declining SHMR and a flattening of the SFMS at the massive end.

Intriguingly, $M_{\rm halo}^{\rm peak} \sim 10^{12}M_{\odot}$ ($M_{\star} \sim 10^{10}M_{\odot}$) also marks a critical transition in black hole growth. Below this threshold, stellar feedback (e.g., supernova-driven winds) dominates and suppresses central black hole accretion. Above this mass, however, compaction events can trigger rapid black hole growth through inflows of cold gas, fueling central starbursts and dense bulge formation \citep{2023MNRAS.522.4515L}. These processes build stellar mass in the galaxy core without significantly increasing galaxy sizes, naturally producing the flattened slope of the high-mass size--mass relation.

At low redshifts, the occurrence of compaction events is significantly reduced compared to high redshifts \citep{2006MNRAS.368....2D}, while a larger fraction of massive galaxies have already transitioned into the quiescent population. Together, these trends steepen the high-mass slope at low redshift for SFGs. By $z \sim0.75$, this slope approaches the low-mass value, possibly indicating a convergence of evolutionary modes in the late universe.

Furthermore, above the pivot mass, halo gas accretion is expected to transition to a hot-dominated mode, reducing the efficiency of cold gas supply and progressively suppressing star formation \citep{2006MNRAS.368....2D}. This suppression can be maintained or enhanced by AGN feedback associated with rapidly growing black holes \citep{2006MNRAS.365...11C}. Together, these processes drive quenching and transform compact, massive SFGs into QGs, implying that cSFGs are the high-redshift progenitors of the compact quiescent population observed in the local universe \citep{2018MNRAS.474.3976G}. 

Consistent with this evolutionary scenario, the pivot mass inferred for QGs is slightly higher than that for SFGs. The pivot mass of QGs increases with redshift, broadly consistent with theoretical expectations that the transitional halo mass from the cold to hot accretion regime becomes higher at earlier cosmic times \citep{2006MNRAS.368....2D,2022A&A...661L...7D}. We will further discuss this in a forthcoming paper (L. Chen. et al. in prep). This characteristic mass is also broadly consistent with the division between low- and high-mass QGs found in, e.g., \citet{2024ApJ...967L..23C} and \citet{Hamadouche2025MNRAS}, further highlighting the distinct physical processes governing the quenching of low- and high-mass galaxies.

\subsection{Implications on the formation pathways of massive QGs}

Previous findings that massive SFGs and QGs follow distinct size distributions pose a serious challenge for physical transformation scenarios: mergers between two extended disks are unlikely to produce compact spheroids unless the progenitors are already compact (and gas-rich) \citep{Wuyts2010}. As one possible resolution, earlier studies identified a population of cSFGs and suggested that they may represent a transitional phase linking massive SFGs to compact QGs through compaction-driven evolution \citep[e.g.][]{2015ApJ...813...23V, 2017ApJ...840...47B, 2016MNRAS.457.2790T, 2023MNRAS.522.4515L,Jiang2025ApJ}. However, due to the limited sensitivity, spatial resolution and wavelength coverage of \textit{HST}, a complete census of cSFGs has been difficult to obtain at $z \gtrsim 3$ before JWST. In fact, even at $z\sim2.5$, a significant population of cSFGs may have been missed in deep HST/WFC3 imaging but could only be unveiled by JWST \citep{2019Natur.572..211W,2025ApJ...993L..49S}. With \textit{JWST}, we are now able to identify a complete sample of cSFGs with reliable structural measurements across a wide redshift range. This allows us to comprehensively study their number density and the overall role in massive galaxy formation and evolution. 

\begin{figure*}
 \begin{center}
  \includegraphics[width=\linewidth]{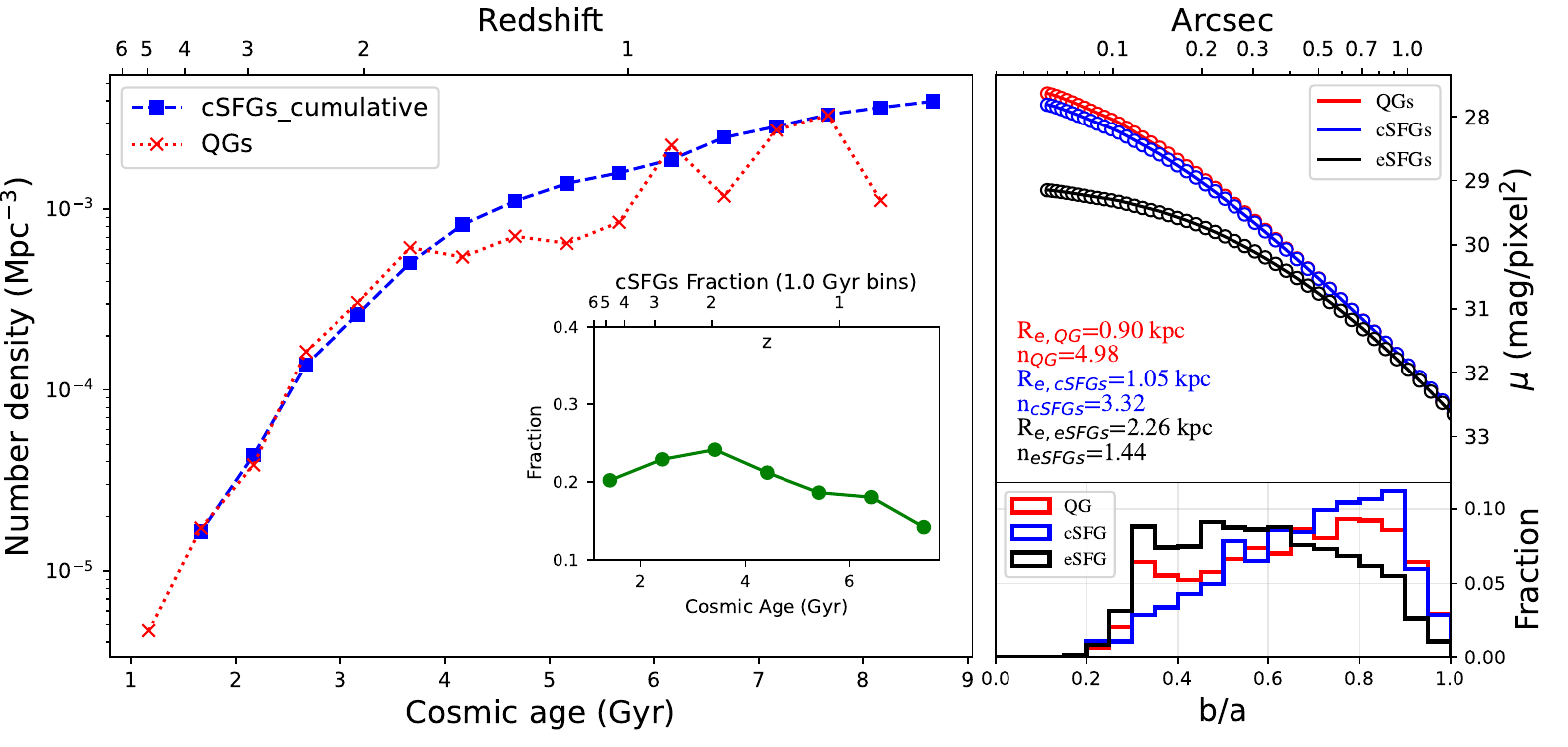}
 \end{center}
\caption{{\it Left}: Evolution of the number density of cSFGs (blue dashed line with square markers) and QGs (red dotted line with “\(\times\)” markers) as a function of cosmic age in 0.5 Gyr bins. The cSFGs curve represents the cumulative number density, obtained by summing the densities from this bin and all preceding (i.e., at younger cosmic ages) bins, while the QGs curve shows the differential number density per bin. The lower axis denotes cosmic age in Gyr, the upper axis gives the equivalent redshift, and all densities are in units of Mpc\(^{-3}\). The inset shows the fraction of cSFGs relative to the total SFG population as a function of cosmic time. {\it Right-top}: Surface brightness profiles (in mag pixel$^{-2}$) as a function of radius for QGs (red), cSFGs (blue), and eSFGs (black). For each population, the data are shown as open circles. The corresponding best-fit single S\'{e}rsic model profiles are shown as solid lines, while the best-fit $n$ and $R_e$ are shown in the bottom-left panel. All profiles are derived from stacked F200W images of galaxies at $2.2 < z < 4.0$, using stellar-mass-matched samples ($10^{10.0}M_{\odot} < M_{\star} < 10^{10.8}M_{\odot}$; $\Delta \log M_{\star} < 0.1$ dex). The typical background level corresponds to a surface brightness of $\sim 34$ mag pixel$^{-2}$.{\it Right-bottom}: the axis ratio (b/a) distribution of QGs (red), cSFGs (blue), and eSFGs (black).}\label{fig:cSFGs}
\end{figure*}

The prevalence of a significant population of massive cSFGs offers an alternative pathway to form compact QGs, which can naturally evolve into compact QGs after exhausting their gas reservoirs, as reflected by their concentrated surface brightness profiles, high $b/a$ and high S\'{e}rsic indices (Figures~\ref{fig:cSFGs} and~\ref{fig:sersic}). Next, we explore whether the number density of these compact SFGs is sufficient to account for that of compact QGs observed across cosmic time.

Assuming that all cSFGs quench within their individual depletion times, we estimate the cumulative number of QGs formed from cSFGs at a given cosmic time. We first compute the gas depletion time for each cSFG using Equation (5) in \citet{2018ApJ...853..179T}, with the parameters set to \(A_t = 0.9\), \(B_t = -0.62\), \(C_t = -0.44\), \(D_t = E_t = 0\), and \(\delta_{\rm MS} = 1\). Specifically, for each redshift corresponding to a data point in the left panel of Figure~\ref{fig:cSFGs}, we count all cSFGs whose predicted gas depletion times imply that they would have quenched prior to that epoch. Dividing by the comoving volume yields the predicted cumulative number density of QGs formed from cSFGs as a function of redshift.

In the left panel of Figure~\ref{fig:cSFGs}, we compare this predicted cumulative number density with the observed number density of QGs over \(0.5 < z < 6.0\). The two curves closely track each other, particularly at $z \gtrsim 2$. The close match between the number density evolution of cSFGs and QGs indicates that the pathway through a cSFG phase should represent a dominant channel for the transformation from SFGs to compact QGs, at least at $z \gtrsim 2$.

\section{Conclusion}

In this work, we establish the size--mass relation for field galaxies over $0.5 < z < 6.0$ using deep \textit{JWST}/NIRCam and MIRI imaging, complemented by multi-wavelength data from other telescopes across a total area of $\sim 300\,\text{arcmin}^2$. The inclusion of MIRI data notably enhances the reliability of photometric redshifts and stellar mass estimates, allowing for robust measurements of galaxy sizes across the full redshift range. We find that neither star-forming galaxies (SFGs) nor quiescent galaxies (QGs) are well described by a single power-law model. The size distribution of SFGs is better characterized by a broken power-law model, while that of QGs is more accurately captured by a mixture power-law model.

Our main results are summarized as follows:
\begin{enumerate}
    \item {\it{A broken power law for SFGs.}} 
    We find that a broken power-law model provides a statistically better description of the size distributions of SFGs than a single power-law model. This model reveals an approximately stable pivot mass at $M_{\star} \sim 10^{10}\,M_\odot$ in the size--mass relation. The low-mass slope is consistent with previous single power-law fits, while the high-mass slope is significantly shallower and becomes progressively flatter toward higher redshifts. These results suggest two distinct size-growth modes, with massive SFGs appearing to decouple from halo growth.

    \item {\it{A mixture model for QGs.}} 
    The size--mass relation of QGs is better described by a two-component mixture model. The transition mass increases with redshift, from $M_{\rm p} \sim 10^{10.0}\,M_\odot$ at $z = 0.75$ to $M_{\rm p} \sim 10^{10.5}\,M_\odot$ at $z = 2.6$, slightly higher than that of SFGs. High-mass QGs are likely direct descendants of massive cSFGs, for which feedback from massive black holes likely plays a major role in both their quenching and maintenance \citep{2024taowang}. On the other hand, the formation of low-mass QGs is probably more related to environmental effects \citep{2010yjpeng, 2025arXiv251012235Y}.

    \item {\it{Compact star-forming galaxies (cSFGs) as quenching progenitors.}} 
    Above the pivot stellar mass of the SFG size--mass relation, we identify a population of compact, massive SFGs whose sizes, number densities, and surface brightness profiles closely resemble those of QGs. The evolution of their number densities indicates that the quenching of these cSFGs can account for the observed buildup of massive QGs at $z \gtrsim 2$, suggesting that compaction, rather than major mergers of extended disks, dominates the formation of massive quiescent galaxies at high redshift.
\end{enumerate}

These results reveal mass-dependent size distributions for QGs and SFGs, which suggest different quenching mechanisms (for QGs) and modes of size growth (for SFGs) at different stellar masses across cosmic time. Future spectroscopic observations with JWST and more detailed morphological analyses (e.g., bulge-disk decomposition) will be essential to reveal the different pathways and underlying physics of size growth and their connection with quenching for galaxies at different redshifts.

\begin{acknowledgments}
We thank the anonymous referee for their constructive comments, which helped improve the clarity and quality of this paper.
This work was supported by National Natural Science Foundation of China (Grant No.12525302 and 12141301), Basic Research Program of Jiangsu (Grant No. BK20250001), National Key R\&D Program of China (Grant no. 2023YFA1605600), the Fundamental Research Funds for the Central Universities with Grant no.KG202502, Scientific Research Innovation Capability Support Project for Young Faculty (Project No. ZYGXQNJSKYCXNLZCXM-P3), and the China Manned Space Program with grant no. CMS-CSST-2025-A04. X. Ding acknowledges support from National Natural Science Foundation of China (Grant No.12573017).

All the \textit{JWST} data presented in this article were obtained from the Mikulski Archive for Space Telescopes (MAST) at the Space Telescope Science Institute. The specific observations analyzed can be accessed via \dataset[DOI:10.17909/xcqt-gw25]{https://doi.org/10.17909/xcqt-gw25}.
\end{acknowledgments}
\appendix
\twocolumngrid
\renewcommand{\thefigure}{\thesection\arabic{figure}}
\renewcommand{\theHfigure}{A\arabic{figure}}

\renewcommand{\thetable}{\thesection\arabic{table}}
\renewcommand{\theHtable}{A\arabic{table}}

\section{effective radius and S\'{e}rsic index recovery test for galight}\label{appendix:galight} 
\setcounter{figure}{0}

\begin{figure*}
 \begin{center}
 \includegraphics[width=\linewidth]{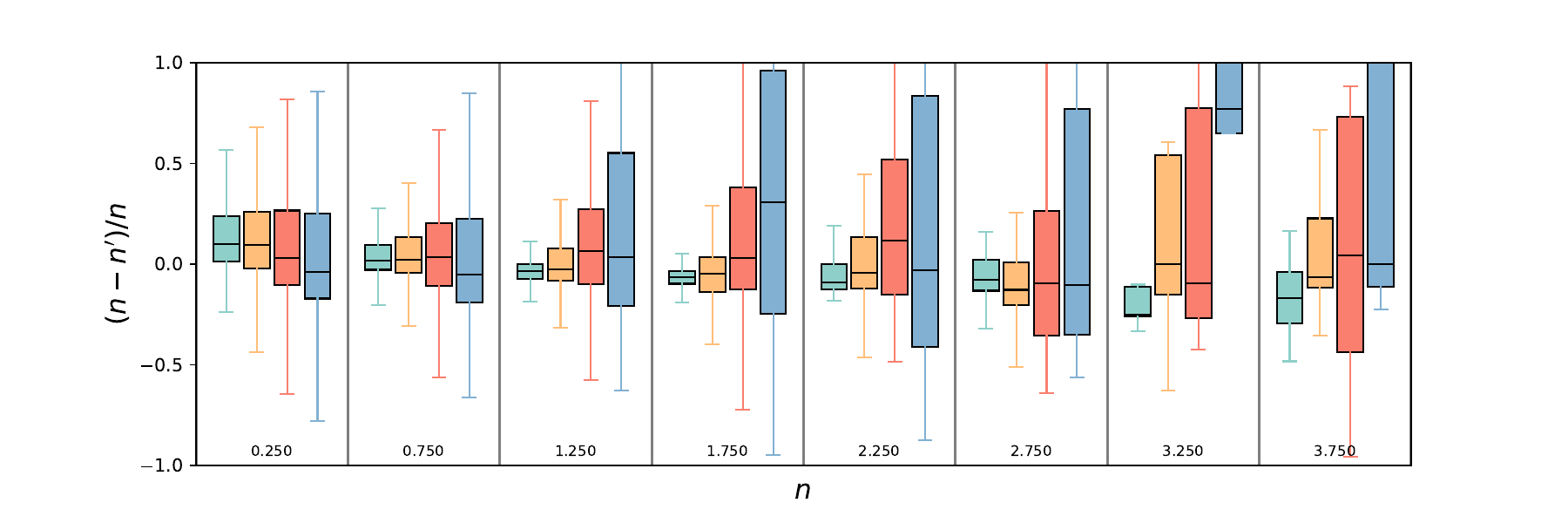}
 \hfill
 \includegraphics[width=\linewidth]{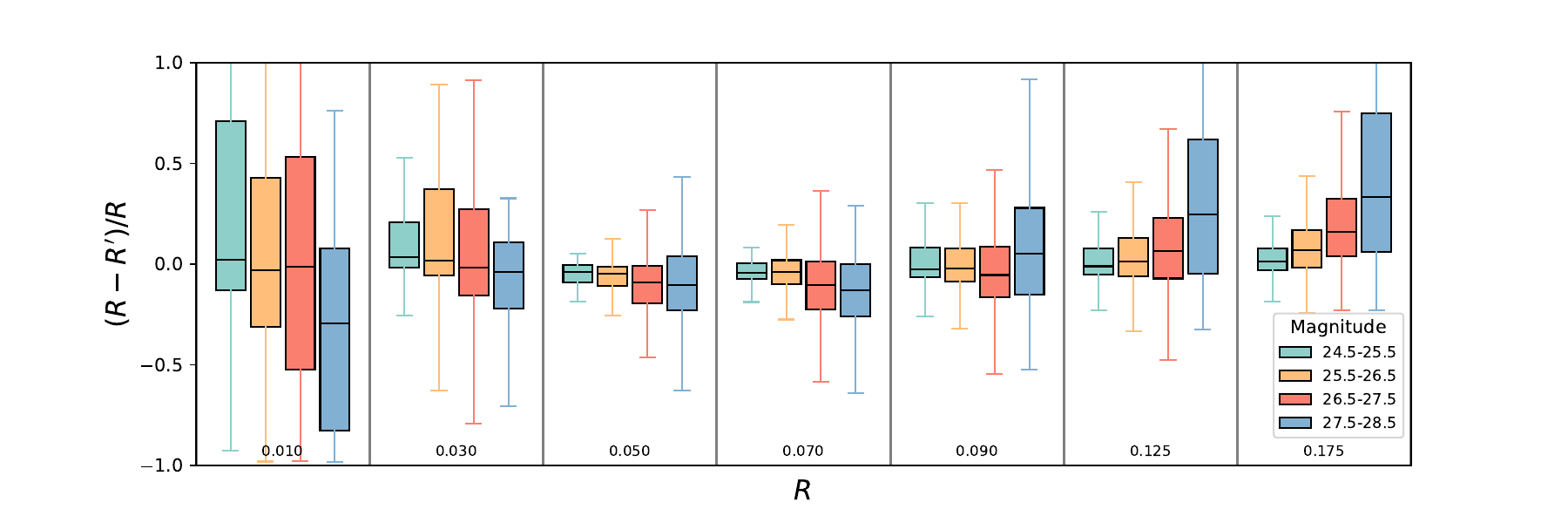}
 \end{center}
\caption{{\it Top}: The relative deviation of the estimated S\'{e}rsic index ($n'$) from the true S\'{e}rsic index ($n$) as a function of $n'$, defined as $(n-n')/n$. The vertical grid lines separate $n'$ bins. Boxes show median (central horizontal line), interquartile range (IQR; box edges), and the maximum/minimum within $1.5 \times \mathrm{IQR}$ (whiskers). {\it Bottom}: The relative deviation of the estimated $R'$ from the true $R$ as a function of $R'$, defined as $(R-R')/R$. Outliers beyond $1.5 \times \mathrm{IQR}$ ($\sim 1.5\%$) are excluded for clarity in all plots.}\label{fig:err_galight} 
\end{figure*}

In order to test the recovery capacity of effective radius and S\'{e}rsic index of \texttt{galight}, we first randomly choose a clean area in CANDELS-COSMOS field with the shallowest observations across all the PRIMER program as background image. We then randomly choose 12569 galaxy models derived from \texttt{galight} 2D image decomposition with $R_e < 0.20 \,\text{arcsec}$ and magnitudes > 24.5. Each galaxy model is located in the background image, and fitted by \texttt{galight} using the same methods as Section~\ref{sec:methods}. The relative deviation of the estimated value $n'$ ($R'$) from the true value $n$ ($R$) is shown in Figure~\ref{fig:err_galight}. We find \texttt{galight} is able to recover $R_e$ with less than $\pm 10\%$ fractional error for galaxies with $R_e > 0.05 \,\text{arcsec}$ and magnitudes < 27.5. The recovery of $n$ is much more challenging than that of $R_e$, but as shown in the bottom panel of Figure~\ref{fig:err_galight}, the errors remain acceptable for galaxies with magnitudes < 27.5, particularly since most of our star-forming galaxies (SFGs) have $n < 1.5$. Based on these tests, we limit all galaxies to $R_e > 0.05 \,\text{arcsec}$ (larger than half of the FWHM of the PSF in each band) and magnitudes < 27.5.

\section{compare with previous work based on HST}\label{appendix:compare} 
\setcounter{figure}{0}
While several recent JWST-based studies and simulations have reported a flattened size--mass relation for massive SFGs at high redshifts, this trend was less clearly established in earlier studies primarily based on HST data. Through detailed re-analysis of the \citet{2014ApJ...788...28V} catalog, we identify two possible resolution-related limitations inherent to HST imaging: (1) The limited angular resolution of HST may lead to either failed structural modeling of intrinsically compact galaxies or systematic overestimation of $R_e$. These galaxies may be removed due to unreliable results or retained with overestimated $R_e$, and both scenarios impact the high-mass end of the size–mass relation, given the increasing fraction of such compact, massive galaxies, and (2) The existence of bright neighboring galaxies can bias $R_e$ measurements for adjacent compact sources through surface brightness contamination. Close galaxies or merging systems may also be erroneously classified as single extended objects, inflating measured sizes. 

\begin{figure*}
\begin{center}
\includegraphics[width=\linewidth]{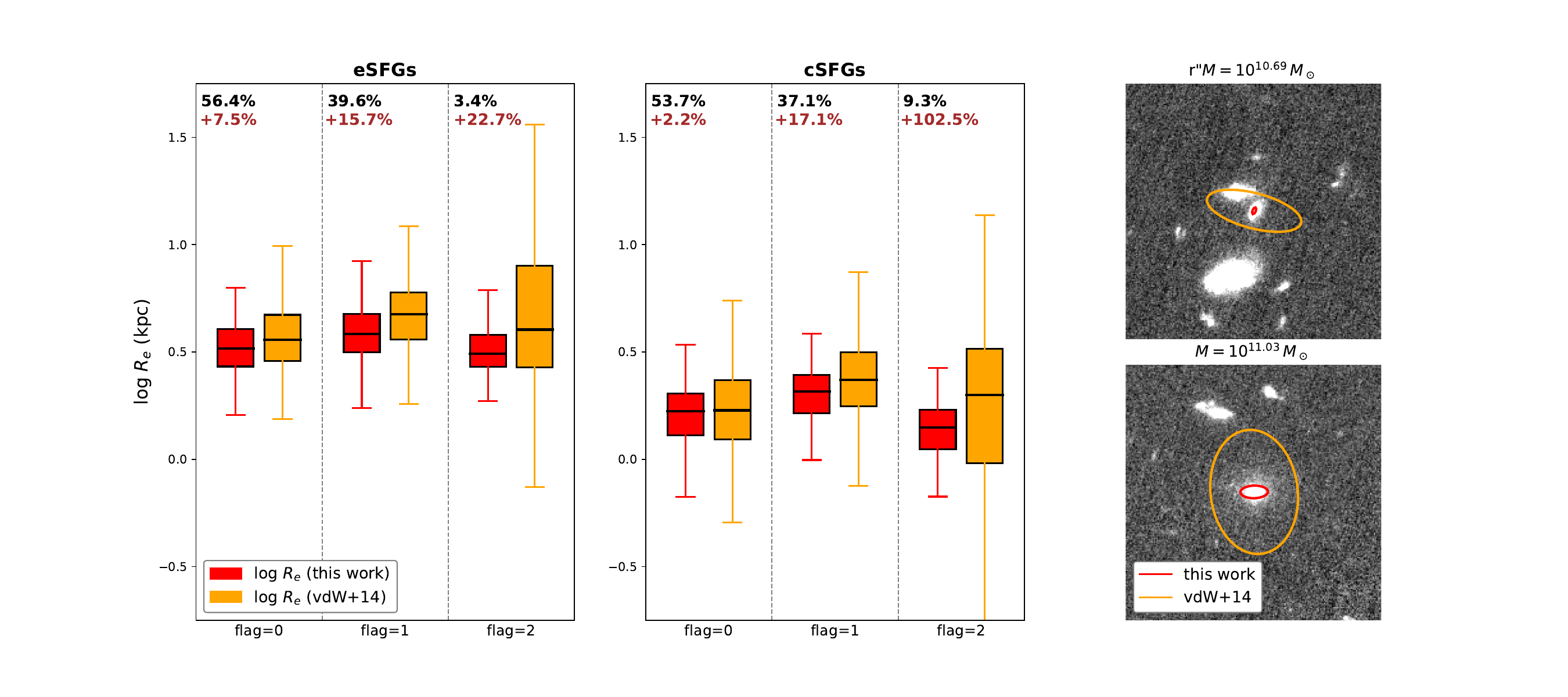}
\end{center}
\caption{
Comparison between galaxy size measurements derived from \textit{HST} and \textit{JWST} imaging.
{\it{Left}}: Size distributions of eSFGs. Red and orange box plots show measurements from this work and from \citet{2014ApJ...788...28V}, respectively. Vertical grid lines indicate different fitting-quality flags defined in \citet{2014ApJ...788...28V}. Each box shows the median (horizontal line), interquartile range (IQR; box edges), and the maximum/minimum within $1.5 \times \mathrm{IQR}$ (whiskers). Outliers beyond this range are excluded for clarity. Red numbers indicate the fraction of galaxies with each flag relative to the total sample. Orange numbers show the percentage by which the median size from \citet{2014ApJ...788...28V} exceeds the median size measured in this work.
{\it{Middle}}: Same as the left panel, but for cSFGs.
{\it{Right}}: Examples of individual galaxies, showing size measurements from both \textit{HST} and \textit{JWST} images. Orange ellipses represent the effective isophotal fits from \citet{2014ApJ...788...28V}, and red ellipses show results from our own fitting.
}\label{fig:comparison} 
\end{figure*}

JWST’s high resolution significantly mitigates limitations in previous size measurements by (1) enabling reliable detection and structural modeling of compact galaxies, and (2) resolving blended systems while reducing contamination, as demonstrated by the comparative HST/JWST analysis in Figure~\ref{fig:comparison}. We cross-match our galaxy sample with that fitted by \citet{2014ApJ...788...28V}, and show the distribution of size differences in the left and middle panels of Figure~\ref{fig:comparison}. As the fitting-quality flag increases (indicating decreasing reliability; see \citealt{2012ApJS..203...24V} for details), the discrepancy between our measurements and those from \textit{HST} becomes more pronounced. This effect is more significant for cSFGs than for eSFGs at the same flag. The right panel presents two representative examples where galaxy sizes derived from HST imaging are clearly overestimated, while JWST-based measurements provide more credible estimates. These results highlight how the improved resolution of JWST enables more robust identification of the broken power-law behavior in the size--mass relation.

\section{The size--mass relation of UVJ-selected samples}
\label{appendix:uvj}
\setcounter{figure}{0}
\setcounter{table}{0}
In addition to the sSFR-based classification of QGs described in Equation~\ref{ssfr_criteria}, we also adopt an alternative criterion based on the UVJ colors following \citet{Williams2009ApJ}. In this approach, QGs are defined as galaxies satisfying
\begin{equation}
\begin{aligned}
V - J &< 1.6 \\
U - V &> 0.88 (V - J) + 0.69 \\
U - V &> 1.3
\end{aligned}
\label{equation:uvj}
\end{equation}
Based on UVJ colors, our sample contains 61573 SFGs and 3776 QGs.

The resulting size--mass relation derived using this UVJ-based classification for SFGs is presented in Figure~\ref{fig:uvj} and Table~\ref{table:uvj} below. We find that the parameters obtained with this method are nearly identical to those derived using the sSFR criterion in the main text. The differences are negligible and do not affect our main conclusions.

\begin{figure*}
 \begin{center}
  \includegraphics[width=\linewidth]{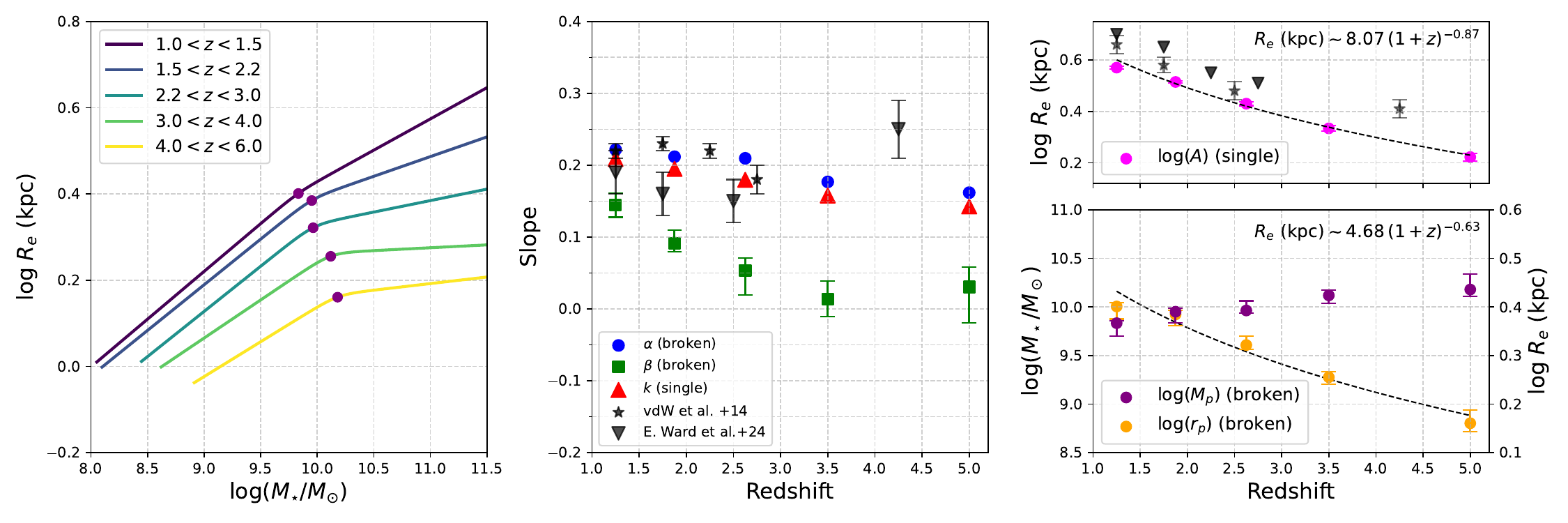}
 \end{center}
\caption{Redshift evolution of the size-mass relations and associated parameters for UVJ-selected SFGs at $1.0 < z < 6.0$ (same symbols as in Figure~\ref{fig:error})}\label{fig:uvj} 

\end{figure*}

\begin{deluxetable*}{lcccccc}
\tablecaption{Best-fit results of the size--mass relations for UVJ-selected SFGs of the form $R_e/{\rm kpc} = r_p(M_{\star}/M_p)^{\alpha}\left[0.5\{1 + (M_{\star}/M_p)^{\delta}\}\right]^{(\beta-\alpha)/\delta}$ (Equation~\ref{eq:broken_sfg}).}\label{table:uvj} 
\tablehead{
  \colhead{Redshift} & \colhead{$\alpha$} & \colhead{$\beta$} & \colhead{$\log(M_{\rm p})$} &
  \colhead{$\log(r_{\rm p})$} & \colhead{$\sigma\log(R_{\rm eff})$} & \colhead{$\Delta BIC$}
}
\startdata
$0.5<z<1.0$ & $0.24_{-0.01}^{+0.01}$ & $0.19_{-0.03}^{+0.01}$ & $9.54_{-0.13}^{+0.39}$ & $0.43_{-0.03}^{+0.09}$ & $0.20_{-0.01}^{+0.01}$ & -2.84 \\
$1.0<z<1.5$ & $0.22_{-0.01}^{+0.01}$ & $0.15_{-0.02}^{+0.02}$ & $9.90_{-0.13}^{+0.03}$ & $0.42_{-0.03}^{+0.01}$ & $0.20_{-0.01}^{+0.01}$ & -0.45 \\
$1.5<z<2.2$ & $0.21_{-0.01}^{+0.01}$ & $0.10_{-0.01}^{+0.02}$ & $10.00_{-0.12}^{+0.03}$ & $0.39_{-0.02}^{+0.01}$ & $0.20_{-0.01}^{+0.01}$ & 44.43 \\
$2.2<z<3.0$ & $0.21_{-0.01}^{+0.01}$ & $0.05_{-0.03}^{+0.02}$ & $10.10_{-0.02}^{+0.10}$ & $0.35_{-0.01}^{+0.02}$ & $0.19_{-0.01}^{+0.01}$ & 25.43 \\
$3.0<z<4.0$ & $0.18_{-0.01}^{+0.01}$ & $0.01_{-0.02}^{+0.03}$ & $10.09_{-0.08}^{+0.05}$ & $0.26_{-0.01}^{+0.01}$ & $0.19_{-0.01}^{+0.01}$ & 6.58 \\
$4.0<z<6.0$ & $0.16_{-0.01}^{+0.01}$ & $0.03_{-0.05}^{+0.03}$ & $10.18_{-0.07}^{+0.16}$ & $0.17_{-0.02}^{+0.03}$ & $0.20_{-0.01}^{+0.01}$ & -7.22 \\
\enddata
\end{deluxetable*}

\section{\texorpdfstring{Size--mass relation at $2.2 < \lowercase{z} < 3.0$ for different samples}{Size--mass relation at 2.2 < z < 3.0 for different samples}}\label{appendix:cosmos_error}
\setcounter{figure}{0}
The differences among the three size--mass relations can be primarily attributed to the presence of a prominent large-scale overdensity in the COSMOS field at $z \simeq 2.45$–$2.65$, corresponding to a massive proto-supercluster. In such overdense environments, massive star-forming galaxies tend to be systematically more compact than their counterparts in the field. As a result, including COSMOS galaxies in this redshift interval leads to a flatter high-mass end slope of the broken power-law relation and a shift of the pivot mass towards higher stellar masses. 
\begin{figure}
 \begin{center}
  \includegraphics[width=\linewidth]{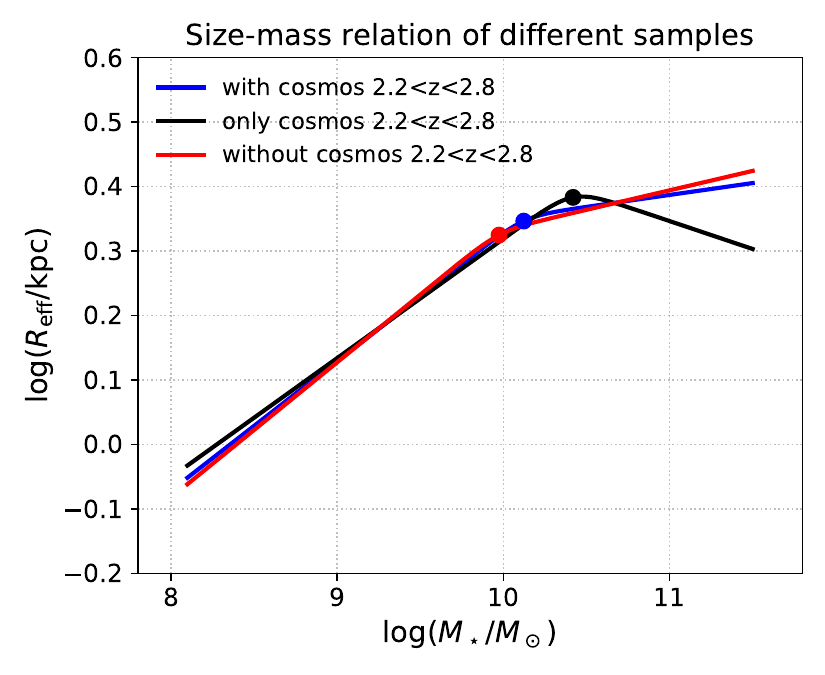}
 \end{center}
\caption{
Stellar size--mass relation of galaxies at $2.2 < z < 3.0$. The solid curves show the best-fitting broken power-law relations obtained from three different samples: the sample including COSMOS galaxies at $2.2 < z < 2.8$ (blue), the COSMOS-only subsample at $2.2 < z < 2.8$ (black), and the sample excluding COSMOS galaxies in this redshift interval (red). The filled circles mark the pivot mass ($\log M_{\mathrm{p}}$) of each fit.
}\label{fig:cosmos_error} 
\end{figure}

\section{Supplementary Figures and Tables}\label{appendix:supplementary}
In this Appendix, we present additional figures and tables that support the main results. Figure \ref{appendix:fig_qg} shows the size–mass relation fitted by a broken power-law model for both SFGs and QGs, and the associated parameters for QGs are shown in Table \ref{appendix:table_qg}. Figure \ref{fig:sersic} shows the distributions of Sérsic indices for SFGs. Figure \ref{fig:stack_image} shows the stack image of cSFGs and eSFGs, along with their Sérsic indices and effective radii.

\setcounter{figure}{0}
\setcounter{table}{0}

\begin{figure*}
\centering
\includegraphics[width=\textwidth]{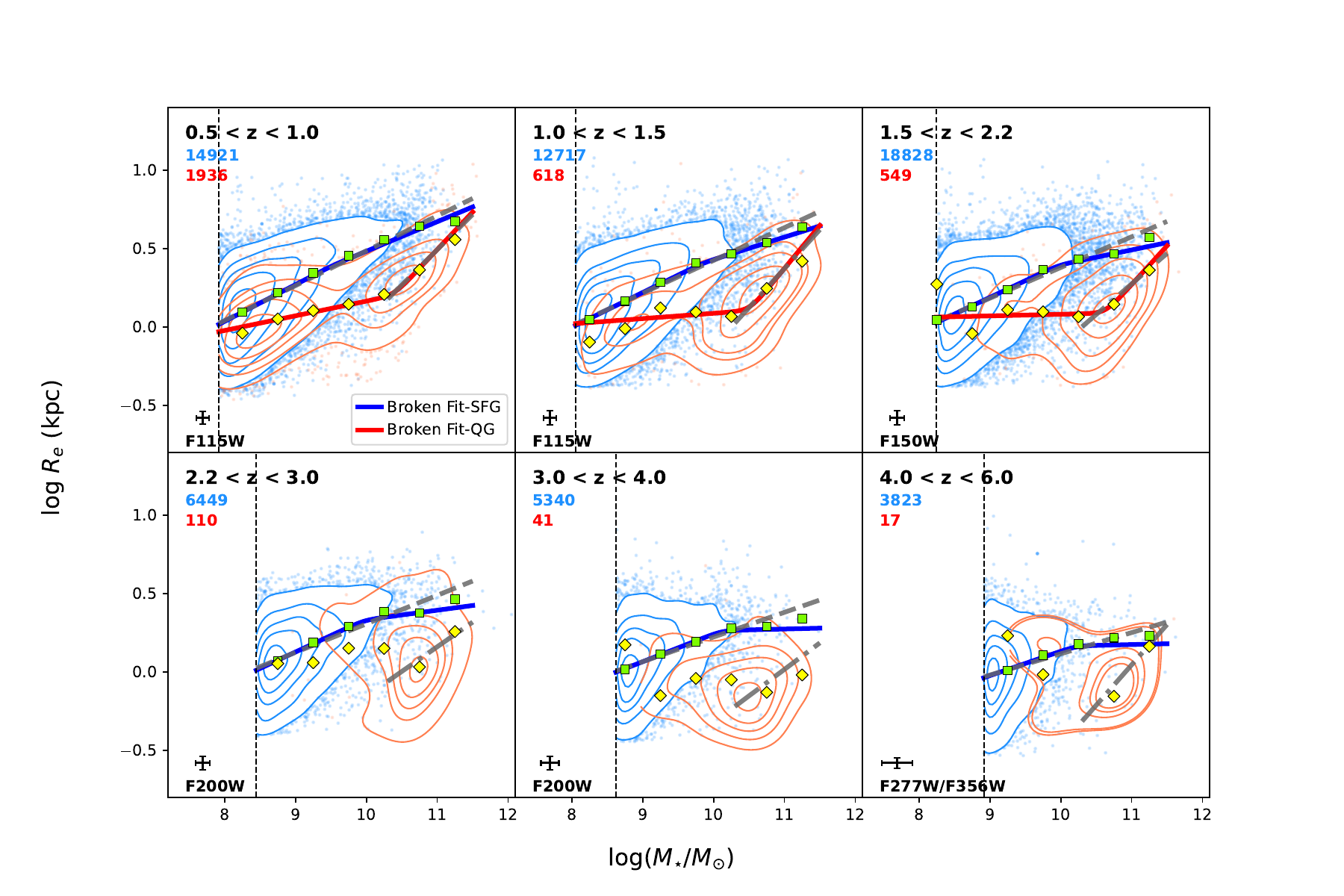}
\caption{Size--mass relations fitted by a broken power-law model for SFGs at $0.5 < z < 6.0$ and for QGs at $0.5 < z < 2.2$ (same symbols as in Figure~\ref{fig:broken}). At $z > 2.2$, the lack of low-mass QGs makes the broken power-law fit become unconstrained and therefore not applicable.}
\label{appendix:fig_qg}
\end{figure*}

\begin{deluxetable*}{lccccccc}
\tablecaption{Best-fit results of the size--mass relations for QGs of the form $R_e/{\rm kpc} = r_p(M_{\star}/M_p)^{\alpha}\left[0.5\{1 + (M_{\star}/M_p)^{\delta}\}\right]^{(\beta-\alpha)/\delta}$ (Equation~\ref{eq:broken_sfg}).\label{appendix:table_qg}}
\tablehead{
  \colhead{Redshift} & \colhead{$\alpha$} & \colhead{$\beta$} & \colhead{$\log(M_{\rm p})$} &
  \colhead{$\log(r_{\rm p})$} & \colhead{$\sigma\log(R_{\rm eff})$}
}
\startdata
$0.5<z<1.0$ & $0.09_{-0.01}^{+0.01}$ & $0.48_{-0.04}^{+0.02}$ & $10.39_{-0.11}^{+0.03}$ & $0.22_{-0.02}^{+0.01}$ & $0.18_{-0.01}^{+0.01}$ \\
$1.0<z<1.5$ & $0.03_{-0.01}^{+0.01}$ & $0.55_{-0.04}^{+0.06}$ & $10.51_{-0.03}^{+0.09}$ & $0.13_{-0.01}^{+0.02}$ & $0.19_{-0.01}^{+0.01}$ \\
$1.5<z<2.2$ & $0.01_{-0.03}^{+0.02}$ & $0.50_{-0.07}^{+0.09}$ & $10.63_{-0.13}^{+0.09}$ & $0.11_{-0.03}^{+0.02}$ & $0.19_{-0.01}^{+0.01}$ \\
\enddata
\end{deluxetable*}

\begin{figure*}
 \begin{center}
  \includegraphics[width=\linewidth]{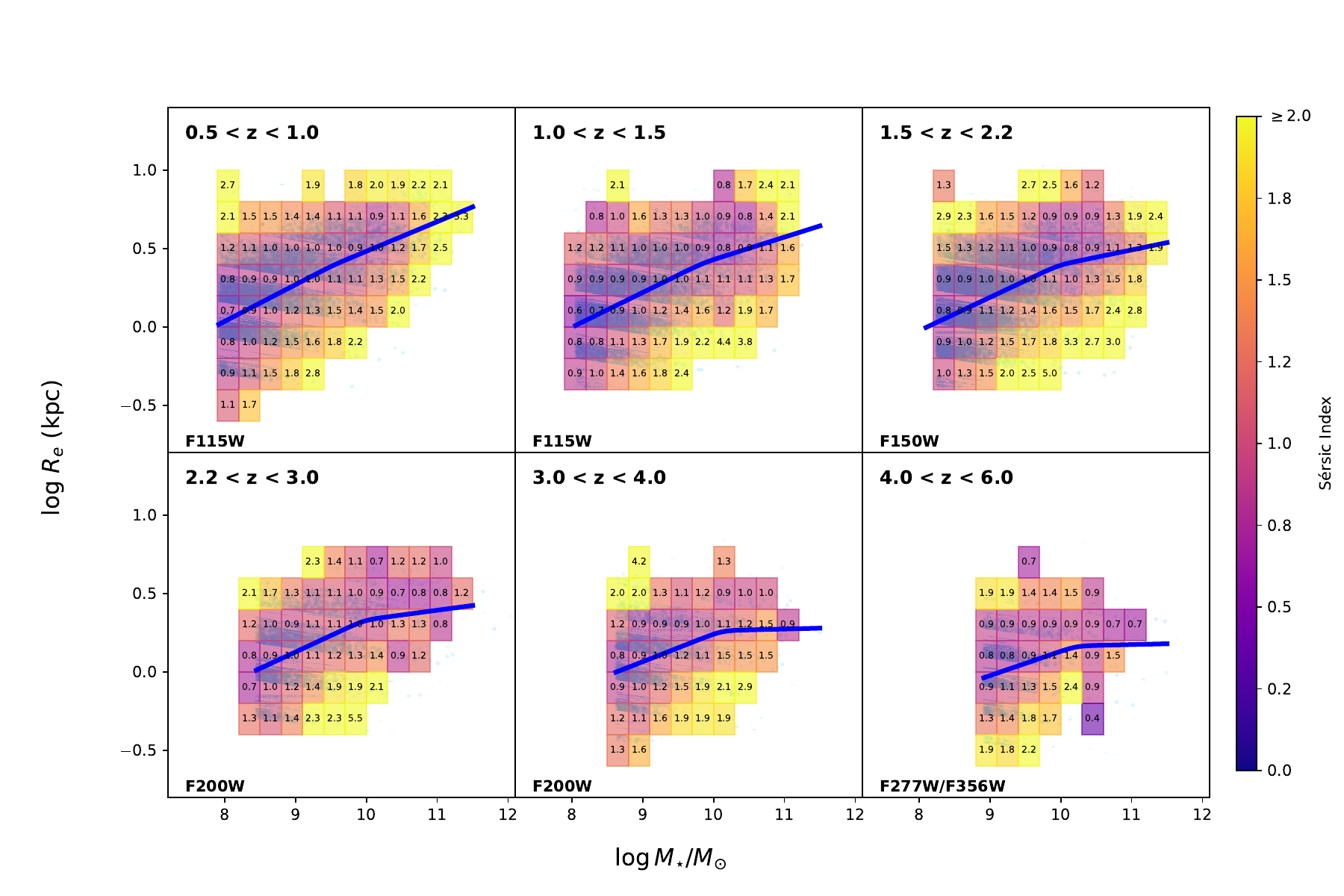}
 \end{center}
\caption{The distributions of S\'{e}rsic index ($n$) for SFGs at $0.5 < z < 6.0$. The numerical values overlaid on each grid cell indicate the median $n$ within that region, and the background color also represents the median $n$. Blue lines  show the size--mass relations fitted by a broken power-law model for SFGs.}\label{fig:sersic} 
\end{figure*}

\begin{figure*}
 \begin{center}
  \includegraphics[width=\linewidth]{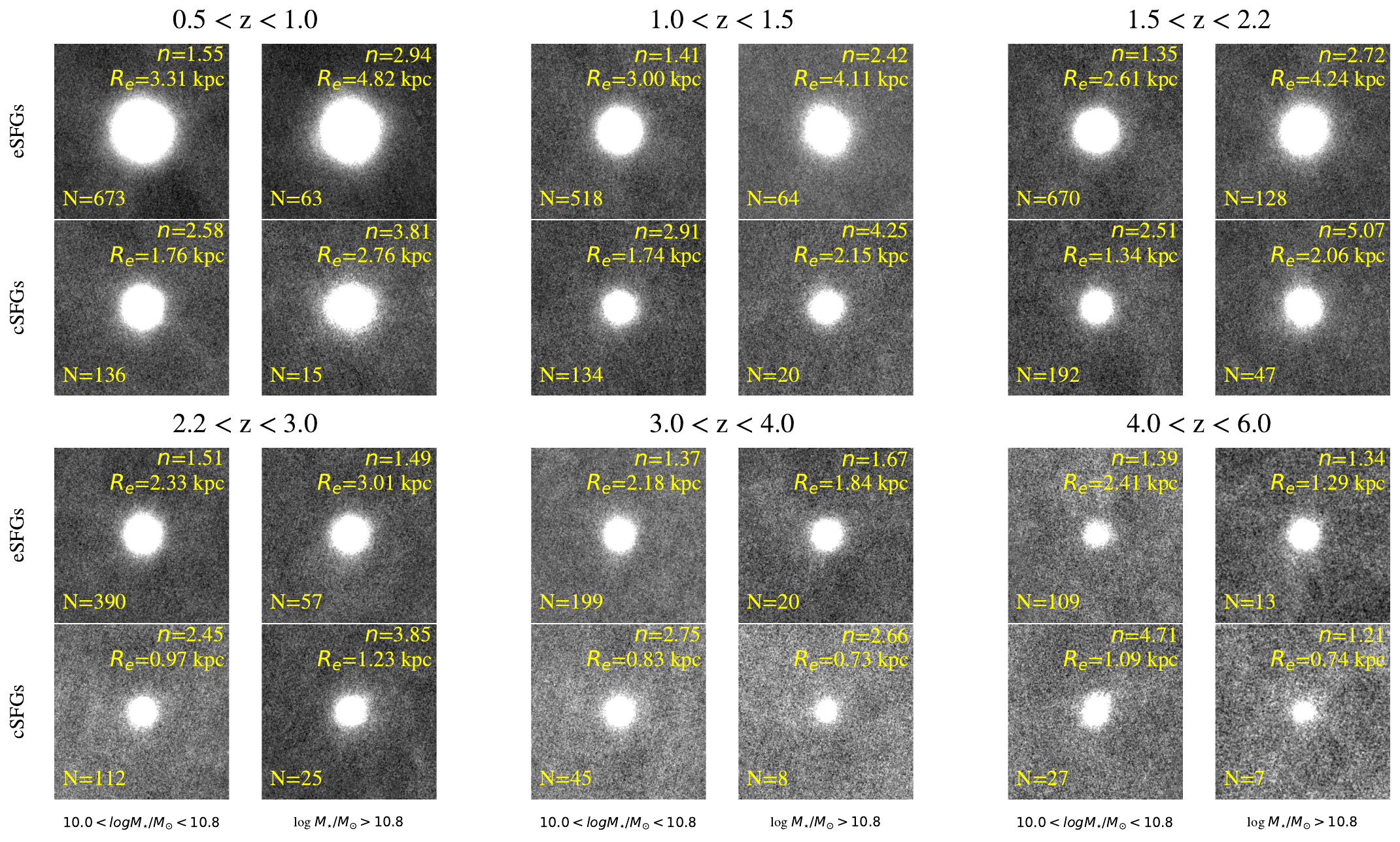}
 \end{center}
\caption{Stacked images of all eSFGs and cSFGs by median value in different stellar mass and redshift bins, using the same band as Figure~\ref{fig:broken}. S\'{e}rsic Index ($n$), effective radius ($R_e$) and the number of galaxies ($N$) are marked on each subplot.}\label{fig:stack_image} 
\end{figure*}

\bibliography{main}{}
\bibliographystyle{aasjournalv7}
\end{document}